# TOWARDS THE AB INITIO BASED THEORY OF PHASE TRANSFORMATIONS IN IRON AND STEEL

**I. K. Razumov[1,2], Yu. N. Gornostyrev[1,2], M. I. Katsnelson[3]**

[1]*Micheev Institute of Metal Physics, UB of RAS, 18 S. Kovalevskaya st., Ekaterinburg, 620990, Russia*
[2]*Institute of Quantum Materials Science, Ural Hi-Tech Park, 5 Konstruktorov st, Ekaterinburg, 620072, Russia*
[3]*Radboud University, Institute for Molecules and Materials, Heyendaalseweg 135, Nijmegen, 6525 AJ, Netherlands*
*e-mail: rik@imp.uran.ru*

Despite of the appearance of numerous new materials, the iron based alloys and steels continue to play an essential role in modern technology. The properties of a steel are determined by its structural state (ferrite, cementite, pearlite, bainite, martensite, and their combination) that is formed under thermal treatment as a result of the shear lattice reconstruction γ (fcc) → α (bcc) and carbon diffusion redistribution. We present a review on a recent progress in the development of a quantitative theory of the phase transformations and microstructure formation in steel that is based on an ab initio parameterization of the Ginzburg-Landau free energy functional. The results of computer modeling describe the regular change of transformation scenario under cooling from ferritic (nucleation and diffusion-controlled growth of the α phase) to martensitic (the shear lattice instability γ → α). It has been shown that the increase in short-range magnetic order with decreasing the temperature plays a key role in the change of transformation scenarios. Phase-field modeling in the framework of a discussed approach demonstrates the typical transformation patterns.

## CONTENTS

## 1. INTRODUCTION

Despite a broad distribution of numerous new materials, steel known from ancient times remains the main structural material of our civilization [1], due to high availability of its main components (Fe and C) and diversity of properties reached by a realization of various (meso)structural states [2–6]. One can control the structural state of steel due to a rich phase diagram of iron with several structural transformations at cooling from moderately high temperatures ( δ → γ → α); the presence of carbon adds carbide phases, cementite $Fe_3C$ being the most important one.

Development of the phase transformations in steel includes two main types of physical processes, namely the crystal lattice reconstruction and redistribution of carbon between the phases. Depending on the rates of these processes and the morphology of decomposition products, metallurgists separate several main types of the transformations, namely, ferrite, pearlite, bainite, and martensite formation, which follow one another with decreasing temperature. All these transformations (except the martensitic one) involve both shear and diffusion mechanisms, their relative importance is changed with the temperature increase. The interplay of different types of transformations determines the diversity of properties of steel and therefore it is of a crucial importance for our understanding of metallurgical processes.

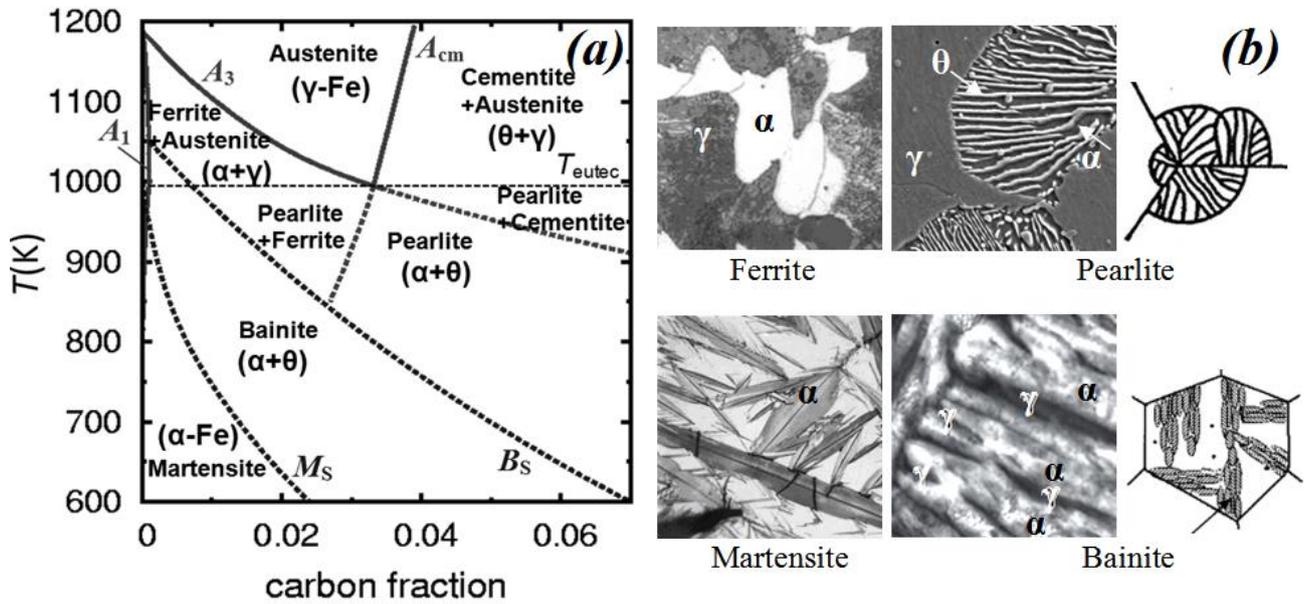

Fig.1. *(a)* Schematic transformations diagram and *(b)* main scenarios of phase transformations in steel. The lines $A_1$, $A_3$ and $A_{cm}$ are the boundaries of two-phase regions $\alpha + \gamma$ and $\gamma + \theta$, as well their metastable extensions below the eutectoid temperature $T_{eutec}$ [16, 17]; $B_S$ and $M_S$ are lines of start of bainitic and martensitic transformations, respectively [4, 18].

Regardless of the great practical significance and comprehensive experimental study, the mechanisms of phase transformations in steel are not fully understood. Firstly, there is still no commonly accepted quantitative theory that could describe the change of the transformation mechanism with temperature increase from martensitic (the lattice instability $\gamma \rightarrow \alpha$ over the entire volume) to ferritic (diffusion-controlled nucleation and growth of precipitates of $\alpha$-Fe). Secondly, the properties of steel are due to the precipitates morphology, understanding of which requires the development of quite a complicated kinetics theory of phase transformations, which takes into account simultaneously the lattice degree of freedom and carbon diffusion.

Based on state-of-the-art first-principle calculations [7, 8] and combining them with the previous models [9–11], we have recently proposed a consistent model of phase transformations in steel [12, 13] that includes a generalized Ginzburg-Landau functional with ab initio parameterization, and nonlinear elasticity equations for the shear transformation and diffusion equation for the carbon concentration. In the framework of this model it was shown that the main factor determining scenarios of the phase transformations in steel is the magnetic state of Fe and its temperature dependence. The constructed curves of the start of ferrite, bainite, and martensite transformations ($A_3$, $T_0$, $M_S$) coincided with the experimentally known ones with good accuracy, and the phase field simulations reproduced the typical transformation patterns. In Ref. [14] this model was generalized, with taking into account the cementite formation, and it was shown that the pearlite colony can emerge by an autocatalytic mechanism at overcooling below the critical temperature.

Here we review the earlier obtained and new results in the framework of this model. In comparison with previous publications, we consider in more detail the results of phase field simulation of transformation kinetics. Also, we discuss the effect of external magnetic field on the curves of transformations diagram.

## 2. PHASE TRANSFORMATIONS AND MICROSTRUCTURE FORMATION IN STEEL

Fig.1a presents the experimental transformation diagram of the Fe-C system, and Fig.1b shows the typical microstructures arising during these transformations. The boundaries of two-phase regions "austenite-ferrite" ($A_1$, $A_3$) and "austenite–cementite" ($A_{cm}$), as well their metastable extensions, - are constructed according to data [15–17]. The lines of the start of the bainitic ($B_S$) and martensitic ($M_S$) transformations is drawn following results in Ref. [18, 19]. Also, the eutectoid temperature $T_{eutec}$ (~1000K) is indicated.

At high temperatures ($T > A_3$, $T > A_{cm}$) the fcc crystal structure of iron ($\gamma$-Fe, austenite) with homogenous carbon distribution is stable. Small overcooling below $A_3$, results in diffusion-controlled precipitation of ferrite (α-Fe, almost pure bcc iron) and the precipitation of cementite (orthorhombic θ phase with 25% at. of carbon) takes place below $A_{cm}$. There are several morphological types of ferrite [20]; allotriomorphic ferrite is usually located at the grain boundaries, whereas the needle crystals of Widmanstätten and acicular ferrite form in the bulk.

If both conditions $T<A_3$ and $T<A_{cm}$ are fulfilled simultaneously, austenite usually decomposes into alternating lamellae of ferrite and cementite, and the interlamellar spacing decreases with overcooling temperature, $\lambda \sim 1/(T - T_{eutec})$. The resulting regular dispersed structure is known as pearlite. The kinetics of pearlite transformation includes the autocatalytic generation of new lamellae, – usually on the grain boundary, – and the growth of the lamellar colony into the bulk [6].

Overcooling below the temperature $B_S$ results in a bainitic transformation, which develops by autocatalytic nucleation and growth of successive sub-units [4]. In the case of upper bainite (which forms in the temperature interval 800K–670K) the ferrite platelets with the same crystallographic orientation are separated by cementite precipitates. In the case of lower bainite, which is formed at a lower temperature, the individual ferrite subunits contain the small ε-carbide precipitates (usually transforming into cementite in the later stages) in addition to the inter-platelet cementite laths. Presumably, a crucial role in the start of bainite transformation is played by a temperature of paraequilibrium $T_0$ where the free energies of the α and γ phases with the same carbon concentration become equal. Temperature $T_0$ was introduced in Ref. [21] as a pre-condition for the start of bainite transformation by displacive mechanism (cooperative displacements of iron atoms). It is assumed in [4, 21] that the diffusion is slower than the shear transformation and therefore there is no redistribution of carbon between the α and γ phases during the growth of α phase plates.

At deeper overcooling (below the start temperature of martensitic transformation, $M_S$) the diffusionless γ→α lattice reconstruction occurs by displacive mechanism, and cementite does not appear. Herewith the twinned structure of martensitic plates compensates for the elastic stresses accompanying the lattice reconstruction. Experimental evidence of the existence of two types of martensite, namely, isothermal (scenario of nucleation and growth of the colony of α-Fe plates with different orientation) and athermal (scenario of spontaneous lattice instability development over the entire volume simultaneously) – were given in [22–25], wherein the first scenario is realized at a higher temperature.

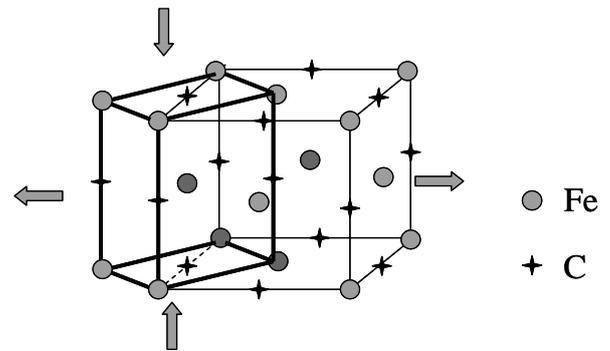

Fig.2. Lattice γ → α reconstruction due to Bain deformation.

## 3. CURRENT UNDERSTANDING OF PHASE TRANSFORMATIONS IN IRON AND STEEL

As it accepted now [4], the displacive mechanism of transformation plays essential role in realization not only the martensitic and the bainitic transformation, but in the formation of high temperature structural state, such as Widmanstätten and acicular ferrite, as well. Thus, understanding of physical mechanisms of the lattice instability of fcc (γ) iron is essential part of our general view on the phase transformations in steel. Two possible mechanisms of γ→α lattice reconstruction were proposed, which correspond to the Bain (tetragonal distortion) [26] and Kurdjumov-Sachs (two shear) [27] deformations schemes. Despite the fact that the Kurdjumov-Sachs scheme is better in corresponding to experiment, the Bain deformation (see Fig.2) usually is considered in the most of theoretical approaches. Reason for this is due to the fact that Bain scheme gives a simplest transformation way that captures, however, important transformation features. Based on the Bain transformation scheme, in Refs. [28, 29] there has been proposed a phenomenological model, being further developed in Refs. [9, 11], which describes the main features of the martensitic transformation, including the formation of coherent systems of twin-like domains.

An overwhelming majority of materials demonstrating the martensitic transformation can be treated as the Hume-Rothery alloys where a particular crystal lattice corresponds to some interval of electron concentration per atom [30]. Electronic mechanisms of the crystal lattice instabilities for these cases are reasonably well understood. They are related to enhanced van Hove singularities in the electron energy spectrum and to the energy gain arising when the Fermi energy lies in a pseudogap [31]. In the new crystal structure, the geometry of the Brillouin zone allows to accommodate all electrons with the essential decrease of the total energy. Since the position of the Fermi energy is determined by the number of electrons per atom, the Hume-Rothery alloys are called also electronic phases. Typically, in these alloys the transformations are close to the second-order phase transitions with a very small hysteresis, the low-temperature phase being more close packed than the high-temperature one. For these alloys a soft-mode picture of phonon spectra is typical [29, 32, 33].

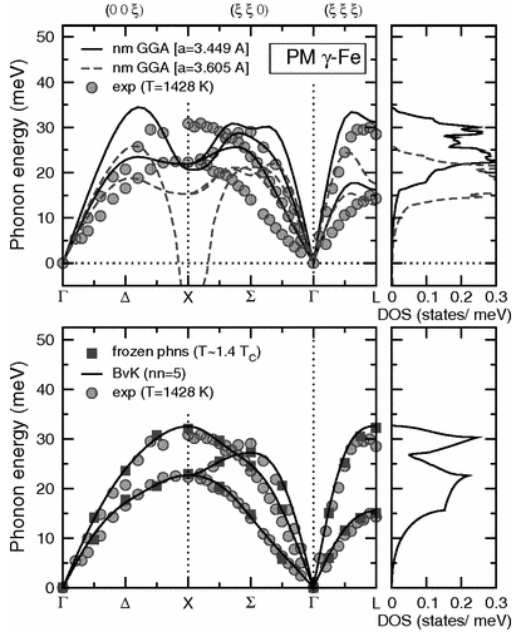

Fig.3. Phonon dispersion curves and corresponding phonon density of states of paramagnetic fcc Fe as calculated within the nonmagnetic GGA (top) and DMFT (bottom) [35]. The DMFT result is further interpolated using a Born-von Kármán model with interactions expanded up to the fifth nearest-neighbor shell. The results are compared with neutron inelastic scattering measurements at 1428K.

However, the iron-based alloys belong to a group of rare materials where the high-temperature phase (fcc, $\gamma$) is close packed and the low-temperature phase (bcc, $\alpha$) is not. Neither experimental data [34] or recent first-principle calculations [35, 36] show soft-mode phonons in fcc Fe above the start temperature of martensitic transformation $M_S$ (see Fig.3). The question whether a soft-mode in phonon spectra appears under cooling, or the mechanism of transition is more complicated than for the Hume-Rothery alloys and cannot be described in terms of individual phonon soft modes, is unclear. The situation looks paradoxical: the $\gamma \rightarrow \alpha$ transformation in iron was historically a prototype of martensitic transitions at all, but this case remains still rather poorly understood, in comparison with many cases discovered later.

Starting from the seminal works by Zener [37], it is commonly accepted that magnetism plays a crucial role in the phase equilibrium of iron and its alloys, including the basic fact that the temperature of the $\gamma \rightarrow \alpha$ transformation in elemental Fe is close to the Curie temperature of $\alpha$-Fe, and bcc iron is stable at low temperatures (see, e.g, [38, 39]). Moreover, the recent first-principles calculations [7, 8, 40] showed that magnetic and lattice degrees of freedom are strongly coupled in $\gamma$-Fe. Therefore, it can be expected, the classical martensitic scenario (through lattice instability over the entire volume) of the $\gamma \rightarrow \alpha$ transformation is realized at overcooling below some temperature where a strong enough short-range ferromagnetic order arises in $\gamma$-Fe (see Section 6 for farther discussion).

The mechanism of bainite transformation that appears for temperature just above $M_S$ remains a subject of debate so far [4, 19]. Two competing theories (diffusion-controlled growth [41–45] and displacive diffusionless nucleation [4, 21, 46]) have been proposed to explain this transformation. This problem has not been solved until now; perhaps the upper bainite formation is a diffusion-controlled process and the lower bainite forms via lattice shearing, as it was assumed in Ref. [47]. Interestingly, in hyper-eutectoid steels the upper bainite is observed even at $T > T_0$ [45, 48], where $T_0$ is the paraequilibrium temperature at which the free energies of the $\alpha$ and $\gamma$ phases with initial carbon concentration are equal [21, 49]; that is in clear disagreement with the displacive model and allows us to consider the upper bainite as a diffusion-controlled nonlamellar eutectoid decomposition product. At the same time, the lower bainite is always formed below $T_0$ [19]. However, in hypo-eutectoid steels the curve of the start of bainitic transformation, $B_S$, is lower than $T_0$, therefore the thermodynamic possibility of shear transformation does not always lead to the formation of upper or lower bainite. Thus, the problem which mechanism controls the bainite transformation is very obscure.

The other sophisticated problem is nucleation and growth of pearlite colonies, which is a particular case of a more general issue of eutectoid decomposition. Transformations of this type are also observed in Zn-Al [50], Cu-Al [51], Au-In [52], Cu-Zn, Al-Mn, Cu-Sn, Cu-Be, etc., and the precipitates morphology (lamellar or globular structure) depends on the type of alloy and the position of alloy parameters on the phase diagram. Although the pearlite transformation (PT) in steel is studied experimentally in detail [53–55], the process of lamellae colony formation remains unclear.

The well-known spinodal decomposition kinetics [56] is not applicable to PT because the mixing energy of carbon in $\gamma$-Fe is positive [57, 58], so that the $\gamma$ phase is stable with respect to small fluctuations of the composition. Thus, more advanced approaches are required for understanding of the PT kinetics. Theoretical studies have been focused on determining the interlamellar spacing and its temperature dependence in a steady state growth of the colony, as well as the problem of stability of the transformation front [21, 59–65]. In Refs. [21, 59] it has been shown that the interlamellar spacing in this case obeys the law $\lambda \sim 1/\Delta T$, where $\Delta T = T - T_{eutec}$. As it was found [61], the interlamellar spacing must ensure a maximum growth rate; thin lamellae dissolute and wide ones split during the growth of colony, thus optimum interlamellar spacing is achieved. Herewith, there was supposed an acceleration of diffusion on the transformation front. The recent results of phase-field simulations [63–65] confirm the necessity of the above assumption. This essential result describes the condition of steady state growth, but it does not concern the problem of nucleation of the pearlite colony, which remains in shadow.

By now, a few important questions are still open. One of them is what phase ($\alpha$ or $\theta$) is appears first or they both form together [66–68]. The question what factors ensure the stability of the front of colony is remained open to discussion [61–65]. The two competing mechanisms of lamellae multiplication by lateral replication [3, 53, 69] and splitting of existing lamellae [70] have been proposed. In addition, the reasons for the transition from lamellar to globular pearlite structure with increasing temperature is a matter of considerable interest [71–75]. There is no theory to explain the appearance of pearlite type colonies under realistic parameters.

Even well-known kinetics of ferrite / cementite precipitation from a supersaturated austenite includes some unresolved problems. In particular, the mechanism of the lattice rearrangement $\gamma \rightarrow \theta$ is under discussion. As is proposed in Ref. [76], the $\gamma \rightarrow \theta$ transformation is realized through an intermediate metastable $\varepsilon$-cementite with hexagonal close-packed (hcp) crystal structure which is closer to $\gamma$-Fe than the orthorhombic $\theta$ phase. The recent ab initio calculations [77] indicate, that lattice $\gamma \rightarrow \theta$ reconstruction can be implemented through a specific Metastable Intermediate Structure (MIS) that develops near the boundary of ferrite plate when the carbon concentration is about 15%at., i.e. far from the stoichiometric composition of cementite. The change of mechanical properties of pearlitic steel after annealing indicates the existence of metastable cementite states in the "fresh" pearlite [3].

In the case of ferrite transformation, the attention of the researchers is drawn to the difference of several morphological forms, polygonal, Widmanstätten, and acicular ferrite (WF, AF) [20, 78, 79]. The polygonal and the Widmanstätten ferrite are realized at a little undercooling (i.e. above the $T_0$ temperature) and, therefore, they both are diffusion-controlled transformations. However, in the first case the lattice coherence on the $\gamma/\alpha$ interface is lost so that elastic stresses are absent, whereas in the second case the elastic stresses relax as a result of twinning of $\alpha$ phase plates. Unlike the two cases mentioned, the acicular ferrite appears below $T_0$ and it grows by the displacive mechanism [4, 20]. Thus, WF and AF can't be described in the framework of simple models with one order parameter. Phase-field simulations of WF formation [80] led to a controversial conclusion that the growth of the WF plates requires high anisotropy of interfacial energy, but the possible role of elastic stresses has been not considered in this work.

Thus, both shear and diffusive scenarios of phase transformations in steels require detailed theoretical study. First, it is necessary to explain the mechanisms responsible for the change of transformation scenarios (ferrite $\rightarrow$ pearlite $\rightarrow$ bainite $\rightarrow$ martensite) with decreasing temperature. Secondly, the precipitates morphology in the decomposition (including the nucleation and growth of the pearlite and bainite colonies, conditions of lamellar or globular pearlite, upper and lower bainite formation, etc.) is a subject of considerable interest. Besides, in some cases (such as bainite or WF transformation) shear and diffusion kinetics should be described together. Discussion of these problems is a matter of the rest part of the present review.

## 4. THEORETICAL APPROACHES TO THERMODYNAMICS AND KINETICS OF PHASE TRANSFORMATIONS IN STEEL

In the framework of a phase-field approach [81] the evolution of microstructure during the martensitic transformation (MT) can be described by the Allen-Cahn equation [9, 11, 82, 83] for a non-conserved order parameter in the capacity of which is the tetragonal deformation $e_t$ is chosen:

$$\frac{\partial e_t}{\partial t} = -\frac{\delta F}{\delta e_t} \qquad (1)$$

$$F[e_t] = \int \left[ f_{el} + \frac{1}{2} k_t (\nabla e_t)^2 \right] d\mathbf{r} , \qquad (2)$$

where $F[e_t]$ is the Ginzburg-Landau free energy functional, $k_t$ is the parameter determining the interface energy, and $f_{el}$ is the nonlinear elastic free energy contribution [84, 85] that is presented as polynomial over $e_t$:

$$f_{el} = f'_{el} + A_2 e_t^2 + A_4 e_t^4 + A_6 e_t^6 \qquad (3)$$

The transformation mechanism in this model switches from the normal type (nucleation and growth) to a martensitic scenario (lattice instability) when the parameter $A_2$ decreases, so one can accept $A_2 = A_{20}(T - T_M)/T_M$, where $T_M$ is the start temperature of MT [10]. However, as was shown in [86], *all* components of the deformation tensor should be taken into account for a proper description of elastic energy at the polymorphic transformation, because they should satisfy a set of Saint-Venant's compatibility conditions $\nabla \times (\nabla \times e)^* = 0$ [87], which can be written in 2D case as:

$$\nabla^2 e_v - \sqrt{8} \partial_{xy} e_s - (\partial_{xx} - \partial_{yy}) e_t = 0, \qquad (4)$$

where $e_t = (\varepsilon_{xx} - \varepsilon_{yy})/\sqrt{2}$ is a tetragonal deformation, $e_v = (\varepsilon_{xx} + \varepsilon_{yy})/\sqrt{2}$ is a dilatation, $e_s = \varepsilon_{xy}$ is a shear (trigonal) deformation, $\varepsilon_{ij}$ are the components of deformations tensor, $\varepsilon_{ij} = (u_{i,j} + u_{j,i} + u_{k,i} u_{k,j})/2$ and $u_{i,j} = \partial u_i / \partial x_j$, and $u_i$ are displacements. Therefore, Eq.(3) should include additional terms:

$$f_{el} = f'_{el} + f_{el}(e_v, e_s) , \\ f_{el}(e_v, e_s) = (A_v e_v^2 + A_s e_s^2)/2 . \qquad (5)$$

The coefficients $A_v$, $A_s$ are expressed in terms of elastic moduli [86], $A_v = C_{11} + C_{12}$, $A_s = 4C_{44}$, and $A_2 = (C_{11} - C_{12})$. As was shown in Refs. [9, 11], by virtue of Saint-Venant's compatibility conditions (4), the Eq.(1) can be converted to a integro-differential form taking into account the effective long-range interactions for the field of order parameter. Due to these long-range effects, the transformation occurs consistently in different microvolumes and is accompanied by the pattern formation that is really characteristic of the MT. Also it was reported on the specific tweed structure that appears as a result of compositional fluctuations and long-range effects at a moderately high temperature.

It should be noted, that an energy-controlled effect of accomodation of elastic stresses that leads to the formation of modulated structures was early considered in Ref. [88]. At the present time the role of long-range interactions in the pattern formation is well known [89, 90, 91] and was discussed many times for very different systems, from stripes in high-temperature superconductors [92–94] to stripe domains in ferromagnetic films [95–97].

The equations of motion for the atomic displacements $\mathbf{u}(\mathbf{r}, t)$ in the form [98]

$$\rho \frac{\partial^2 u_i(\mathbf{r}, t)}{\partial t^2} = \sum_j \frac{\partial \sigma_{ij}(\mathbf{r}, t)}{\partial r_j} , \qquad (6)$$

$$\sigma_{ij}(\mathbf{r}, t) = \frac{\delta F}{\delta \varepsilon_{ij}(\mathbf{r}, t)} ,$$

are more convenient than Eq. (1) for numerical simulation of the transformations kinetics. Here $\rho$ is the density, $\sigma_{ij}(\mathbf{r}, t)$ are the components of elastic stresses, $i, j = \{x, y\}$. The solution of Eq.(6) satisfies the Saint-Venant's compatibility conditions automatically and can contain also the lattice vibrations (lattice temperature). This approach has been used in the simulations of MT in Refs. [10, 99].

First theoretical description of the martensitic type phase transformation in 3D case was proposed in Refs. [28, 29]. This approach is based on the expansion of the Ginzburg-Landau functional over deformations relevant for the lattice transformation and taking into account only the deviatoric components ($e_t, e_{tt}$) of the deformations tensor:

$$f(e_t, e_{tt}) = A(e_t^2 + e_{tt}^2) + \\ + Be_{tt}(e_{tt}^2 - 3e_t^2) + C(e_t^2 + e_{tt}^2)^2 \qquad (7)$$

где $e_t = (\varepsilon_{xx} - \varepsilon_{yy})/\sqrt{2}$, $e_{tt} = (\varepsilon_{xx} + \varepsilon_{yy} - 2\varepsilon_{zz})/\sqrt{6}$. Within this model, the transformation mechanism is changed from "nucleation and grows" to lattice instability development (martensitic scenario) when parameter $B$ decreases.

In Ref. [10] the contributions related to carbon concentration and its interplay with deformations have been added to the free energy, and the system of equations for atomic displacements (6) and the Cahn-Hilliard equation for carbon diffusion [56] was resolved by phase-field simulations for the martensitic and pearlitic scenarios of phase transformations. The equation for the carbon diffusion has a form:

$$\frac{\partial c}{\partial t} = -\nabla I , \qquad I = -\frac{D}{kT} c(1 - c) \nabla \left( \frac{\delta F}{\delta c} \right), \qquad (8)$$

where $c$ is a local carbon concentration, $D$ is a carbon diffusion coefficient. Herewith the free energy functional has a form of:

$$F[e_t, e_v, e_s, c] = \int[f_{el} + f_{ch} + f_{cpl} + \frac{1}{2}k_t(\nabla e_t)^2 + \frac{1}{2}k_c(\nabla c)^2]d\mathbf{r}, \quad (9)$$

where $f_{el}$ is the density of elastic free energy, which includes the term given by Eq. (5) and contribution of concentration expansion, $f_{ch} = v_2 c^2 + v_4 c^4$ is the chemical contribution to the free energy, and the term $f_{cpl} = A_t^{(c)} c^2 e_t^2$ takes into account the coupling between elastic distortions and composition changes. Typical transformation patterns obtained in the framework of this approach are shown in Fig.4. The model [10] is one of the first attempts to take into account the interplay between the diffusive and displacive mechanisms of phase transformation. However, this approach is pure phenomenological and contains assumptions that are incorrect for steels. For example, the coupling contribution $f_{cpl}$ did not include the linear concentration term, although the solution energies of carbon are different in the γ and α phases [100–102]. Besides, the proposed model assumes that mixing energy of carbon in the γ phase is negative in disagreement with experiment [57] and ab initio calculations [58]. Finally, precipitates morphologies of pearlite obtained in the simulations were far from those observed in experiments [3, 53, 69].

It should be noted that the mechanism of colony formation in the system with a positive mixing energy (such as the γ phase) is unknown. The proposed approaches considered mostly the evolution of existing colony of alternating plates of ferrite and cementite placed on the flat grain boundary [62–65]. The model of eutectoid transformation in a system with a symmetric phase diagram was considered in Ref.[62] where the growth of two phase lamellae was observed in the case of equal diffusion coefficients in the different phases (see Fig.5). In a more realistic case the widths of cementite and ferrite lamellae are different, the carbon diffusion coefficient in ferrite is much more larger than this one in cementite and austenite, therefore an assumption of the diffusion acceleration on the colony front is required to provide steady state growth [63, 64].

The problems of early stages of the colony formation and the multiplication of lamellae remain outside of the scope of proposed models. The similar issues exist in the eutectic colonies growth, where the metastable liquid phase decomposes into two new phases at the solidification under a temperature gradient [103–107] or without it [108–110].

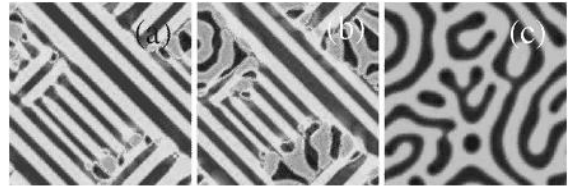

Fig.4. The appearance and evolution of martensitic structure to pearlite-like one in the model taking into account an interplay between diffusive and displacive phase transformations; (a) $t$=4000, (b) 6000, (c) 12000 [10].

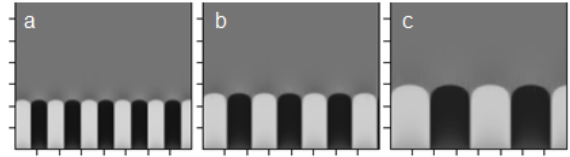

Fig.5. The structure of stationary growing colonies in eutectoid system with a simmetric phase diagram at different temperatures; $T/T_{eutec}$= (a) 0.59, (b) 0.70, (c) 0.82 [62].

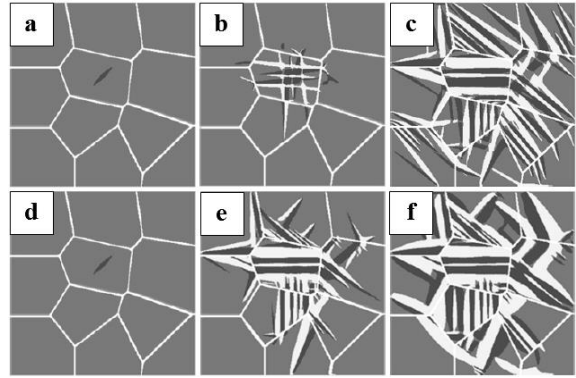

Fig.6. Evolution of martensite in 2D case with only (a)–(c) elastic and (d)–(f) with elasto-plastic deformations; $t$= (a,d) 0, (b,e) 25, (c,f) 100 [111].

As was discussed above, the regular martensite pattern formation is driven by the accommodation of elastic stresses to minimize the energy. In last decade the attention of researchers was attracted to the problem of the plastic accommodation of transformation strains that provides another relaxation channel of elastic energy minimization [111–114]. It was shown that accounting for the plastic relaxation processes results in the possibility of the easier martensitic transformation and a more complex and coarser microstructure (see Fig.6). It should be noted that the essential role of plastic deformation in a phase transformation was early predicted in Ref. [115, 116] where a single ellipsoidal nucleus has been considered. The general phase-field approach including a system of the coupled equations for the order parameters of phase transformation and the mechanics equation for dislocation-assisted plasticity was proposed in Ref. [114].

The main features of the pattern formation in the course of the martensitic-type structural phase transitions proved possible to describe within the framework of the models proposed in Refs. [9–11, 112, etc.]. The scenarios of athermal [10, 86] (lattice instability over the entire volume in the case of rapid quenching), isothermal [9, 11, 117] (autocatalytic generation of martensite plates in the case of holding the steel at a moderate temperature), stress-assisted, and strain-induced [112, 113, 118] MT were discussed. However, it remained unclear how to apply more correctly these model approaches to the real iron and steel.

The general disadvantage of the theoretical approaches considered above is the phenomenological form of free energy density. In particular, the authors do not distinguish the enthalpy and entropy contributions to the free energy density, therefore the microscopic meaning of parameters appears lost and their correct choice is impracticable.

## 5. THE AB INITIO BASED MODEL OF SHEAR-DIFFUSION PHASE TRANSFORMATIONS IN IRON AND STEEL

The consistent model of phase transformations in steel should take into account (1) the lattice reconstruction $\gamma \rightarrow \alpha$ during cooling to the critical temperature (Bain [26] or Kurdjumov-Sachs [27] deformation); (2) Saint-Venant's compatibility conditions [87] for the components of deformations tensor leading to the appearance of effective long-range interactions for the field of the order parameter [9–11]; (3) redistribution of carbon between the phases, including the formation of cementite. Herewith, the Ginzburg-Landau free energy functional should include the magnetic energy contribution.

### 5.1. Ab initio parameterization of Bain transformation path.

The total energy per atom along the Bain deformation path was calculated for both ferromagnetic ordered and paramagnetic (disordered local moment, DLM) states of iron [7, 8]. The difference between energies in these states is related to magnetic exchange energy. The ab initio computational results show that the appearance of ferromagnetic order leads to the change of the preferable crystal structure of Fe from fcc to bcc (see Fig.7). In [8] was also shown that there is strong coupling between the magnetic and lattice subsystems in fcc Fe so that exchange energy drastically changes due to the volume increase or tetragonal distortions (see Fig.8). In addition, the ferromagnetic ordered fcc structure is unstable in respect to fcc → bcc transformation. These results suggest that the martensitic transformation of Fe can appear as a result of lattice instability due to the increase in short-range magnetic order under the cooling.

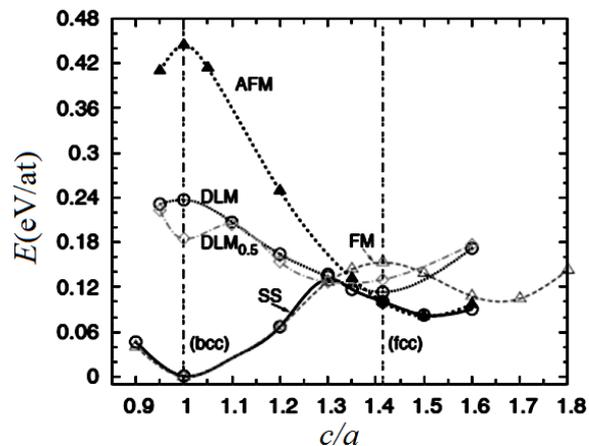

Fig.7. Variation in the total energy per atom along the Bain deformation path for different magnetic states. FM (empty triangles) and AFM (solid triangles) label collinear ferromagnetic and antiferromagnetic structures, SS (empty circles) corresponds to the spin-spiral state, DLM (crosses) belongs to the disordered local moments approximation of paramagnetic state, $DLM_{0.5}$ (empty diamonds) stands for the DLM state with a total magnetic moment equal to half of that for the FM state [7].

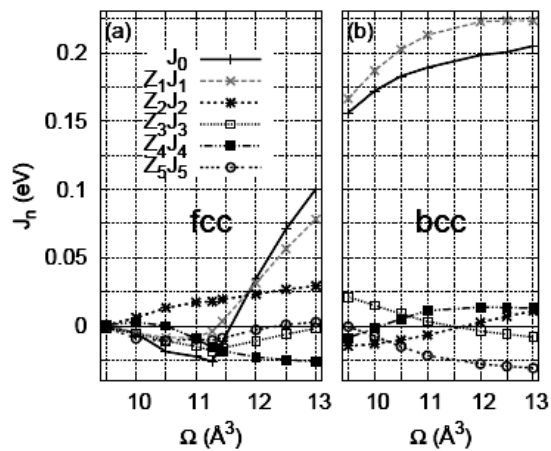

Fig.8. Exchange parameters $J_n$ for n=1,2,3,4,5 and the total exchange parameter $J_0$ in dependence on atomic volume for (a) fcc and (b) bcc Fe [8].

The first-principles computational results allow us to find an explicit expression for the density of free energy for pure Fe, which takes into account both deformations and magnetic degrees of freedom. For this purpose, we represent the magnetic-dependent part of the total internal energy in Heisenberg-like form

$$E = E_{PM}(\hat{\varepsilon}) - \sum_{i<j} J_{i,j}(\hat{\varepsilon}) Q_{ij}(T) , \qquad (10)$$

where $Q_{ij}(T) \equiv <\mathbf{m}_i \cdot \mathbf{m}_j >$ is the correlation function of magnetic moments on sites $i$ and $j$, $E_{PM}$ is the energy of paramagnetic state, and the brackets $<\dots>$ mean the average over an ensemble of magnetic configurations at a given temperature. Assuming that the nearest-neighbor contribution is dominate in exchange interactions, the energy density can be presented as

$$g(\hat{\varepsilon}, T) = g^{PM}(\hat{\varepsilon}) - \tilde{J}(\hat{\varepsilon}) Q(T) , \qquad (11)$$

where $\tilde{J} = m^2 J / \Omega$, $\Omega$ is the volume per atom and $m$ is the magnetic moment, $Q(T) \equiv <\mathbf{m}_0 \cdot \mathbf{m}_1 > / m^2$ is the spin correlation function dependent on temperature; $Q = 0$ stands for the totally disordered paramagnetic (PM) state and $Q = 1$, for the ferromagnetic (FM) ground state. The exchange energy $J(\hat{\varepsilon})$ can be extracted from the computational data [7,8],

$$\tilde{J}(\hat{\varepsilon}) = g^{PM}(\hat{\varepsilon}) - g^{FM}(\hat{\varepsilon}) . \qquad (12)$$

It is assumed here that $J(\hat{\varepsilon})$ depends on the Bain tetragonal deformation $e_t$ only, and the value of dilatation is chosen from the minimum of energy at a given $e_t$.

In order to determine the spin correlation function, the model proposed by Oguchi for the total spin equal to 1/2 [119] was employed as a benchmark in Refs. [12,13]. In this model

$$Q(T) = \frac{(2\mathrm{ch}(h) + 1) - 3e^{-2j}}{(2\mathrm{ch}(h) + 1) + e^{-2j}} , \qquad (13)$$

$$h = h_0 + (z-1)\sigma j , \quad h_0 = \frac{g\beta H_0}{kT} , \quad j = \frac{\tilde{J}}{zkT} , \qquad (14)$$

where $g \approx 2$ is the Lande factor, $\beta$ is the Bohr magneton, $H_0$ is the external magnetic field (if it is present), $\sigma$ is the reduced magnetization determined from the transcendental equation:

$$\sigma = \frac{2\mathrm{sh}(h)}{e^{-2j} + 2\mathrm{ch}(h) + 1} , \qquad (15)$$

The essential advantage of the Oguchi model (compared with the well-known Langevin formula for the magnetization) is the accounting for the short-range magnetic order at $T > T_C$, where $T_C$ is the Curie temperature.

Based on these formulas, in Ref. [13] there was accepted that $Q(T) \sim 1/T$ (without an external magnetic field) for $T > T_C$, and the empirical dependence of magnetization [120] was used for $T < T_C$. It was assumed there that $Q(T_C) \sim 0.4$, according to Ref. [119]. The Curie temperature $T_C$ is related to the exchange parameter as $kT_C(e_t) = \lambda \tilde{J}(e_t) \Omega$, with the numerical factor $\lambda_\alpha = 0.472$ for $\alpha$-Fe; this choice of $\lambda_\alpha$ provides an agreement of the Curie temperature with the experiment, $T_C = 1043K$. The correlator for $\gamma$-Fe is chosen in a similar way, with the Curie temperature $T_C^\gamma \approx 300K$, according to the calculations [8] for the fixed atomic volume $\Omega \approx 12\,Å^3$; $\lambda_\gamma = 0.606$ according to Ref. [121] (see [13] for details). The nonphysical ferromagnetic long-range ordering in $\gamma$-Fe is not essential for our model, but the high-temperature short-range ordering in both $\alpha$ and $\gamma$ phases is important enough for the transformation kinetics.

Assuming that tetragonal deformation is counted from an fcc phase ($e_t=0$ in the $\gamma$ phase and $e_t = 1 - 1/\sqrt{2}$ in the $\alpha$ phase) we consider the order parameter $-1 < \phi < 1$ which is related to the Bain deformation as $\phi = \sqrt{2} / (\sqrt{2} - 1) e_t$. Positive and negative values of $\phi$ correspond to the two possible (mutually orthogonal) directions of the Bain deformation in two-dimensional case.

The energies found from the first-principle calculations [7, 8, 13] for pure iron were approximated by the following polynomials:

$$\tilde{g}^{PM(FM)}(\phi) = g_\gamma^{PM(FM)} +$$
$$+ 2\left( g_\alpha^{PM(FM)} - g_\gamma^{PM(FM)} + \frac{c^{PM(FM)}}{6} \right) \times \qquad (16)$$
$$\times \left( \phi^2 - \frac{\phi^4}{2} \right) + c^{PM(FM)} \left( \frac{\phi^6}{3} - \frac{\phi^4}{2} \right)$$

Its form guarantees an extremum at the points $\phi = 0$ or $\phi = \pm 1$, and parameters $g_{\gamma[\alpha]}^{PM(FM)}$, $c^{PM(FM)}$ were found by fitting to the ab initio computational results (see [13] for the details).

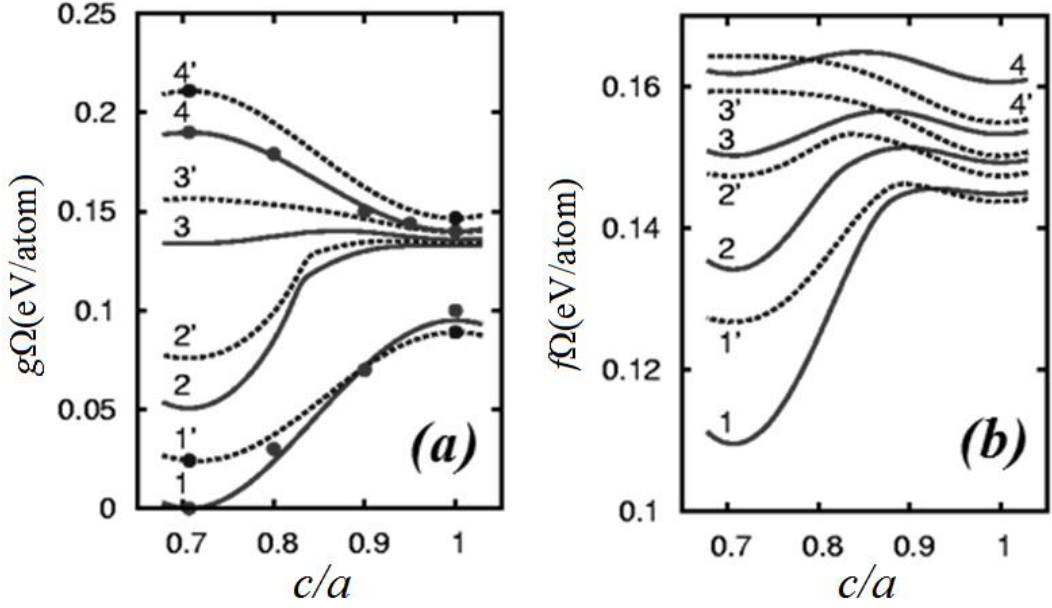

Fig.9. *(a)* Energy resulting from the first-principle calculation for the Bain path at $T$=0K (curves 1,1'), 800K (curves 2,2'), 1400K (curves 3,3') and in paramagnetic states (curves 4,4') and *(b)* the free energy density as a function of tetragonal deformation for the temperatures $T$=600K (curves 1,1'), 800K (2,2'), 1000K (3,3'), 1400K (4,4') found from Eqs. (17). The carbon concentration is $C$ = 0 (1–4) and $C$ = 3at% (1'–4'); circles correspond to ab initio calculated results.

The Bain path energies are modified in the presence of carbon:

$$g^{PM(FM)}(\phi,c) = \tilde{g}^{PM(FM)}(\phi) + \varepsilon_\gamma^{PM(FM)} c +$$
$$+ v_\gamma c^2 / 2 + (1 - f_s(\phi)) \times \qquad (17)$$
$$\times \left[ \left( \varepsilon_\alpha^{PM(FM)} - \varepsilon_\gamma^{PM(FM)} \right) c + (v_\alpha - v_\gamma) c^2 / 2 \right]$$

where function $f_s(\phi)$ have been chosen in the form $f_s(\phi) = (1 - \phi^2)^2$. Thus, in our approximation the dependence of Bain path energy on carbon concentration is reduced to the accounting for the solution energies of carbon, $\varepsilon_{\gamma[\alpha]}^{PM(FM)}$, and the energies of carbon-carbon interactions (mixing energies) $v_{\gamma[\alpha]}$ in the $\gamma$ and $\alpha$ phases, but not in the intermediate states. The solution energies were obtained from ab initio calculations (see [13] for the details) and mixing energies from Refs. [57, 58]. It should be noted that the known estimates of $v_\gamma$ vary widely from 1 to 3 eV/at, i.e. the $\gamma$ phase is to be stable with respect to carbon decomposition. If the changes of carbon concentration are small or moderate (as in the cases of ferrite, martensite, and in the early stages of bainite transformations) the contribution of carbon-carbon interactions can be neglected [13]. However, in the cases of the cementite formation (i.e. in case of pearlite and at later stages of bainite transformations) the carbon concentration increases essentially (up to

$c$=0.25); the carbon-carbon interactions must be taken into account in this case.

Dependences of the energy density $g(c/a)$ on tetragonal distortion calculated according to the formulas (16),(17) are shown in Fig.9a. One can see that $\gamma$-Fe is stable in paramagnetic state but it looses its stability with respect to tetragonal (Bain) deformation when becoming ferromagnetic. Therefore, the classical martensitic scenario (through lattice instability over the entire volume) of the $\gamma \rightarrow \alpha$ transformation can emerge at the overcooling below some temperature where a strong enough short-range ferromagnetic order arises in $\gamma$-Fe. The doping by carbon does not change this important feature. Moreover, carbon decreases the energy of ferromagnetic $\gamma$-Fe, with the solution energy of the order of -0.2 eV per carbon atom. It is not surprising, since carbon creates a strong local ferromagnetic order in PM or AFM $\gamma$-Fe [40]. It is a common wisdom that interstitial impurities (including carbon) always prefer fcc surrounding compared to bcc, just for geometric reasons [122] (the voids are larger in fcc lattice than in bcc with the same density). This is for sure correct, also for carbon in iron and results in a more pronounced effect of carbon addition on energy of $\alpha$-Fe. What is much less trivial is that carbon solubility in $\gamma$-Fe is sensitive to the magnetic state being maximal in ferromagnetic surrounding.

## 5.2. Generalized Ginzburg-Landau functional for the $\gamma - \alpha$ transformation in steel.

Bain path energy is important part to construct a quantitative theory of phase transformations in steel. The Ginzburg-Landau (G.-L.) functional of free energy of iron and steel should include also contributions related to the magnetic, phonon, electron, and carbon configuration entropy, energy gain during formation of cementite, energy of elastic stresses, and interphases energies. The G.-L. functional can be written in the form similar to Eq.(9):

$$F = \int \Big( f(c,e_t,\eta,T) + f_{el}(e_\nu,e_s) + \\ + \frac{k_t}{2}(\nabla e_t)^2 + \frac{k_\eta}{2}(\nabla e_\eta)^2 \Big) d\mathbf{r} , \qquad (18)$$

where $f(c,e_t,\eta,T)$ is local density of free energy depending on the carbon concentration $c$, tetragonal deformation parameter $e_t$, temperature $T$, and order parameter $\eta$ characterizing the transformation of austenite to cementite at the point $\mathbf{r}$; $f_{el}(e_\nu,e_s)$ is the elastic energy determined by Eq. (5); $k_t$ and $k_\eta$ are the parameters determining the width of ferrite or cementite interphase boundary, respectively [86].

Let us first consider the local density of free energy in the absence of cementite. This situation is typical of the ferritic (in low-carbon steel at $T > T_{eutec}$) and martensitic (i.e. below the temperature $M_S$) transformations. Using the Hellmann-Feynman theorem and Eq.(11) one can represent the free-energy density for the elemental Fe as

$$f(e_t,T) = g_{PM} - Ts_0 f_s(e_t) - \int_0^{\tilde{J}} Q(\tilde{J}',T) d\tilde{J}' , \quad (19)$$

where $s_0$ is the high-temperature limit of the entropy difference between the fcc and bcc phases, including phonon contribution; $f_s(e_t)$ is a function provided a gradual switching of the entropy contribution from fcc to bcc ( $f_s(e_t) = 1$ in fcc and $f_s(e_t) = 0$ in bcc) phase. According to existing concepts (see, for example, [123]) the value $s_0$ depends slightly on the temperature at $T > T_D$, where $T_D$ is the Debye temperature (473K in bcc and 324K in fcc phase). It has been chosen such that the start of the transformation determined by the condition $\Delta f(T) = f_\gamma(T) - f_\alpha(T) \equiv 0$ agrees with the experimental value for elemental Fe, $T_0 = 1184$K. This gives us the value $s_0 = -0.19k$, quite close to the experiment [124].

The temperature dependence of the energy $\Delta g(T) = g_\gamma(T) - g_\alpha(T)$ and free energy difference

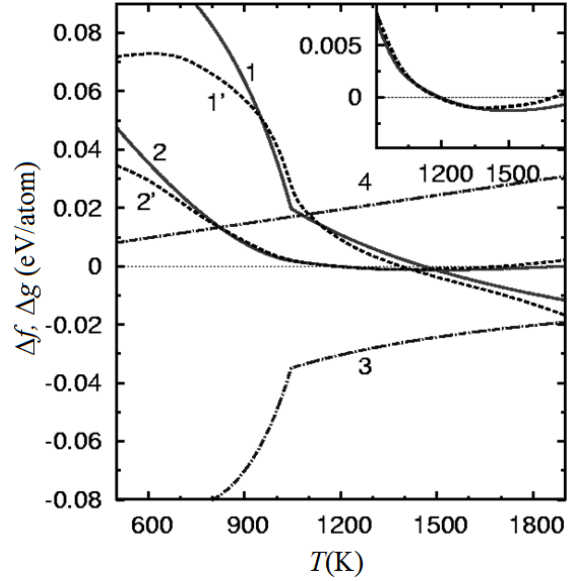

Fig.10. The energy difference $\Delta g(T) = g_\gamma(T) - g_\alpha(T)$ (curve 1) and free energy difference $\Delta f(T) = f_\gamma(T) - f_\alpha(T)$ (curve 2) at the $\gamma \rightarrow \alpha$ transition in elemental iron in comparison with known data (dotted lines 1',2') [124]; contribution of magnetic entropy to the free energy (curve 3) and the contribution from phonon entropy (curve 4).

$\Delta f(T) = f_\gamma(T) - f_\alpha(T)$ for the pure Fe agrees well with the results of CALPHAD [124] within the temperature range 600÷1200K (see Fig.10). Herewith, the magnetic contribution dominates at $T \le T_C$ and is compensated essentially by the phonon contribution at $T > T_C$.

The configurational entropy of carbon is found from the model of ideal solutions, assuming that for $T > 300$K carbon is equally distributed among all three interstitial sublattices in α-Fe, whereas in γ-Fe carbon atoms can occupy only quarter of the interstitial positions [88,102]. As a result, the local density of free energy can be presented as:

$$f(c,e_t,T) = g_{PM} - Ts_0 f_s(e_t) - \int_0^{\tilde{J}} Q(\tilde{J}',T) d\tilde{J}' - \\ - T\Big[ S_\gamma + (S_\alpha - S_\gamma)(1 - f_s(e_t)) \Big] \qquad (20)$$

where $S_{\alpha(\gamma)}$ is the configurational entropy of carbon in the α(γ) phase, $S_\alpha \approx -kc\ln(c/3)$, and $S_\gamma = -k\big[ 4c\ln(4c) + (1-4c)\ln(1-4c) \big]/4$.

Dependences of the local density of free energy on tetragonal distortion, calculated according to the formulas (20), are shown in Fig.9b. It can be seen from a comparison of Fig.9a and Fig.9b that the curves $g(e_t)$ and $f(e_t)$ are qualitatively similar, but they differ in the depth of the minima corresponding to the phases α and γ. In particular, the minimum corresponding to the γ phase exists on the curve of $f(e_t)$ up to the sufficiently low temperature about 400K. It means that lattice reconstruction requires an overcome of some energy barrier at an experimental temperature $M_S$, and one follows to expect that the martensitic transformation occurs by the nucleation and growth of an embryo in this case.

The lattice reconstruction γ→θ leading to cementite formation is another structural transformation, which is controlled by carbon diffusion. Herewith, the order parameter $\eta$ in Eq.(18) describes a preferred trajectory of the transition $\gamma \rightarrow \theta$ including the Metastable Intermediate Structure (MIS) [77]. According to these ideas, the MIS appears in the thin ferromagnetic layer existing near the ferrite plate. The subsequent lattice reconstruction MIS→θ occurs by the cooperative displacements mechanism when the local carbon concentration increases to a threshold value $c$~0.18. Then the θ phase is saturated with carbon to the stoichiometric composition of cementite ($C_{cem}$ =0.25). As a result, the lattice coherence is maintained, whereas the elastic stresses are well compensated at the interface α/θ.

Since the lattice reconstruction is a rather fast process (unlike diffusion), γ→θ can be considered as immediately occurring as soon as the free energies of austenite and cementite become equal. Then, the local carbon concentration is a single order parameter characterizing the cementite, and the density of its free energy [14] can be written as

$$f_\theta(c,T) = f_{\alpha-Fe}(T) + \Delta f_{\alpha\theta}(T) + \\ + \left(f_\theta^{(1)}(c) - f_\theta^{(1)}(c_{cem})\right) + \Delta f_\theta^{bound}(T), \quad (21)$$

where $f_{\alpha-Fe}(T)$ is the free energy of the pure α-Fe, $\Delta f_{\alpha\theta}(T)$ is free energy of formation of cementite from the pure compounds (α-Fe and graphite) known from CALPHAD and ab initio calculations [125, 126], $c_{cem}$ is the stoichiometric composition of cementite ($c_{cem}$=0.25), $f_\theta^{(1)}(c)$ is the concentration dependence of free energy of cementite [127]. The value of $\Delta f_\theta^{bound}$~ -0.02eV/at is a shift of free energy of cementite due to magnetization induced by the adjacent ferrite plate; $\Delta f_\theta^{bound}$=0 if an isolated cementite nucleus is considered.

By using the approach proposed in Ref. [10] the transformations kinetics can be described by the system of coupled equations for the atomic displacements (6) and carbon redistribution (8). The carbon diffusion coefficient we define as:

$$D(c) = \left[D_\gamma + \left(D_\alpha - D_\gamma\right)h(C_{T0} - c)\right] \times \\ \times h(C_{T1} - c) + D_\theta h(c - C_{T1}), \quad (22)$$

where $h(x)$ is a smoothed Heaviside function, $C_{T0}, C_{T1}$ are the carbon concentrations corresponding to the conditions of paraequilibrium, namely, $f_\gamma(c,T) = f_\alpha(c,T)$ and $f_\gamma(c,T) = f_\theta(c,T)$, respectively. Eq. (22) provides that the carbon diffusion coefficient is equal to $D_\alpha, D_\gamma, D_\theta$ in the bulk of the respective phases and it takes the intermediate values $D(c)$ at the interfaces. The ratios of the coefficients $D_\alpha / D_\gamma$, $D_\gamma / D_\theta$ are $10^2$ or $10^3$ [128, 129], thus the simulation with realistic diffusion coefficients is unfeasible, but the qualitative trends can be derived by choosing a reasonably large value of ratios of the diffusion coefficients.

# 6. CONSTRUCTION OF TRANSFORMATIONS DIAGRAM OF STEEL

The model proposed in Refs [12–14], which includes the lattice and magnetic degrees of freedom, allows to construct the transformation diagram of the system Fe-C. This diagram (Fig.11) includes the boundaries of two-phase regions γ/(α+γ), γ/(θ+γ) (lines $A_3$ and $A_{cm}$ respectively, see Fig.1), as well their extensions into metastable region below the eutectoid temperature $T_{eutec}$, and also the lines of instability in respect of the γ→α and γ→θ transitions ($T_0$ and $T_1$, respectively).

The lines $A_3$ and $A_{cm}$ are defined by the equality of the chemical potentials of carbon, and the lines $T_0$ and $T_1$ are determined by the equality of the free energies of corresponding phases at a fixed carbon concentration:

$$\frac{df_\alpha}{dc}(C_{A1}) = \frac{df_\gamma}{dc}(C_{A3}) = \frac{f_\gamma(C_{A3}) - f_\alpha(C_{A1})}{C_{A3} - C_{A1}}, \quad (23)$$

$$\frac{df_\gamma}{dc}(C_{Acm}) = \frac{df_\theta}{dc}(C_{cem}) = \frac{f_\gamma(C_{Acm}) - f_\alpha(C_{cem})}{C_{Acm} - C_{cem}}$$

$$f_\alpha(C_{T0}) = f_\gamma(C_{T0}), \quad f_\gamma(C_{T1}) = f_\theta(C_{T1}), \quad (24)$$

where $f_{\alpha(\gamma,\theta)}$ are the free energy density of the α(γ,θ) phases, wherein $f_{\alpha(\gamma)}$ are determined by the Eq.(20)

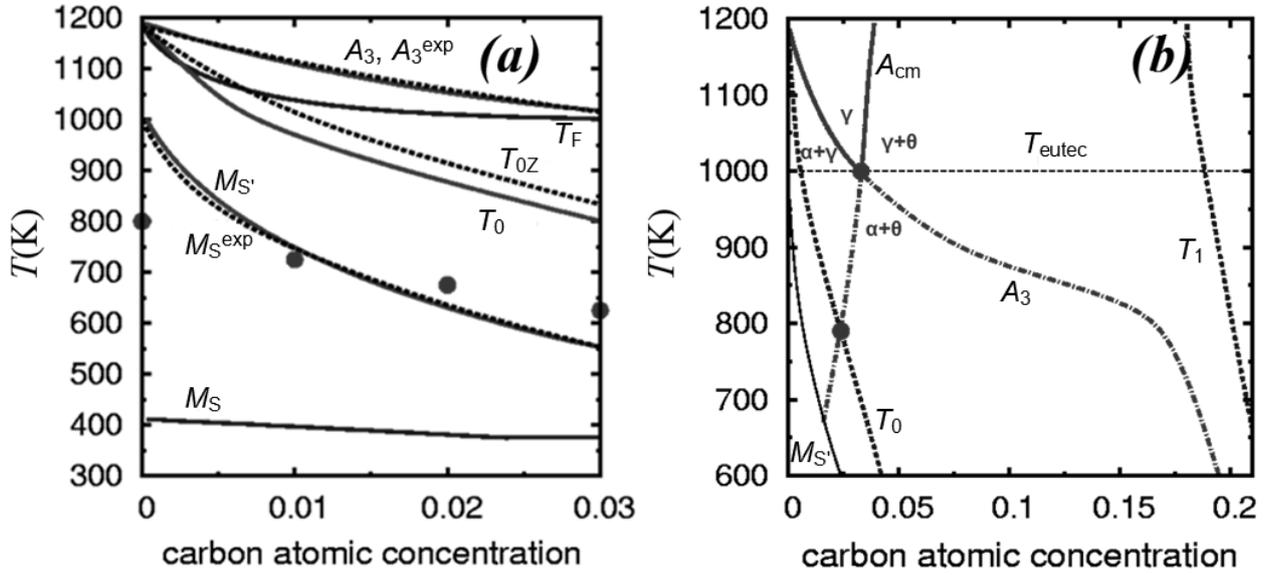

Fig.11. **(a)** Calculated lines (solid) corresponding to the start of ferrite formation, shear nucleation (line of paraequilibrium), and martensitic transformation [13]. $M_S$ and $M_{S'}$ are the start temperatures of the absolute lattice instability and martensitic colonies nucleation. Dashed lines show experimental boundary of two-phase region ($A_3$) [16], experimental paraequilibrium temperature ($T_{0Z}$) [49], and experimental temperature of the start of martensitic transformation ($M_S^{exp}$) [18]. The circles correspond to the start of martensitic transformation in phase-field simulations with thermal fluctuations. **(b)** Transformation diagram taking into account the formation of cementite [14]. The boundaries of two-phase regions $\gamma/(\alpha+\gamma)$, $\gamma/(\theta+\gamma)$ (lines $A_3$ and $A_{cm}$, respectively) with their metastable extensions and the curves of paraequilibrium $\gamma/\alpha$, $\gamma/\theta$ ($T_0$ and $T_1$ respectively) are calculated.

at $e_t = e_t^\gamma = 0$ in the $\gamma$ phase and $e_t = e_t^\alpha = 1 - 1/\sqrt{2}$ in the $\alpha$ phase, respectively, and $f_\theta$ is determined by the Eq.(21) at $\Delta f_\theta^{bound} = 0$. The line $T_F$ was construct by using the condition $f\left(e_t^\gamma, c_0, T\right) = f\left(e_t^\alpha, c = 0, T\right)$, where $c_0$ is the initial (average over the sample) carbon concentration; and the line $M_S$ (the start of martensitic transformation) is defined by the disappearance of the barrier on the Bain deformation path, $\partial^2 f\left(e_t, c, T\right)/\partial e_t^2 = 0$. The line $A_1$ is close to a zero carbon concentration and is not presented here. The transformation diagram at a low carbon concentration (Fig.11a), without the formation of cementite, was constructed in Ref. [13] and at a high carbon concentration (Fig.11b) discussed in Ref. [14].

Let us first consider an expected qualitative picture of the ferritic and martensitic transformation scenarios in low- and medium-carbon steel [13]. At a low enough overcooling below the temperature $A_3$ the ferrite transformation proceeds slowly, since its driving force is small. The nucleation of ferrite as a result of thermal fluctuations is scarcely probable, because the critical size of nucleus (determined by the ratio of surface energy to the bulk energy gain) is very large. This is more likely at the grain boundaries where the chemical potential of carbon is changed. Herewith, the growth rate of ferrite is limited by carbon diffusion in the $\gamma$ phase in this case, because the energy of the $\alpha$ phase without carbon is higher than the energy of the $\gamma$ phase with an initial composition. In the case $T \le T_F$ the ferrite nucleus can grow even if the composition of the $\gamma$ phase remains unchanged and the growth rate of ferrite accelerates essentially.

A further decrease of temperature results in a slowdown of carbon diffusion and enhancement of the transformation driving force. At intermediate temperatures, a crucial role in determining of the start of transformation is played by the temperature of paraequilibrium $T_0$ (see formula (24)), below which the free energy density of the $\alpha$ phase is less than the free energy density of the $\gamma$ phase with the same carbon concentration. Temperature $T_0$ was introduced in Ref. [21] as a pre-condition for the start of bainite transformation. Since diffusion is slower than the shear transformation [4, 21], there is no redistribution of carbon between the $\alpha$ and $\gamma$ phases during the growth of $\alpha$ phase plates. In low-carbon steels the relation $T_{eutec} < T < T_0$ is possible, when the shear lattice reconstruction is realized in the ferrite region of the transformations diagram. The corresponding transformation scenario can be interpreted as acicular ferrite, which is realized by displacive mechanism [20]. As can be seen in Fig.11a, the calculated value of $A_3$ and $T_0$ agrees well with the known experimental quantity $A_3^{exp}$ and $T_{0Z}$ [16,49].

The condition $\partial^2 f\left(e_t,c,T\right)/\partial e_t^2 = 0$, corresponding to the disappearance of the barrier on the Bain deformation path, is attained by quenching of the $\gamma$ phase to the temperature $M_S$ where ferromagnetic short range order in the $\gamma$ phase becomes important. Below this temperature the absolute lattice instability of the $\gamma$ phase should take place, and the obtained martensite can be called as the athermal one. It can be seen that the temperature $M_S$ found in this way is actually lower than the experimental value. However, according to the concept of isothermal martensitic transformation [22–25] the condition of the martensite start may be taken as $f_{barrier}^{\gamma \to \alpha} = \tilde{C}_0 kT$, where parameter $\tilde{C}_0 = 0.04$ is chosen by fitting to the experiment for pure Fe [18]. The temperature $M_{S'}$ determined in this way agrees well with the experiment in a broad interval of carbon concentrations.

The possibility of overcooling of austenite to the liquid nitrogen temperature, with the formation of martensite only during the subsequent heating, was first discovered in Ref. [22]. This observation is clearly points to thermally activated character of the transformation with a rather small activation energy value of about 0.04eV/at [22]. It has been later shown that the isothermal kinetics changes to an athermal one in some Fe-based alloys at overcooling below some critical temperature [2]. The avoiding of discussion of this problem can lead to some kind of misunderstanding. For example, the athermal martensite was in focus of the model used in Ref. [10], whereas the kinetics of the nucleation and growth of isothermal martensite has been investigated in Ref. [11].

The transformations diagram shown in Fig.11b takes in addition into account the formation of cementite (for details of parameterization see Ref. [14]). The line $A_{cm}$ is the boundary of the two-phase region $\gamma + \theta$. The intersection of $A_3$ and $A_{cm}$ lines is the eutectoid point ($c_{eutec}, T_{eutec}$); below the themperature $T_{eutec}$ the pearlite transformation (PT) occurs. In accordance with a traditional point of view, the PT is realized between the $A_3$ and $A_{cm}$ lines extrapolated into a $T < T_{eutec}$ region ("Hultgren extrapolation" [130]), where the simultaneous nucleation of $\alpha$ and $\theta$ from the initial $\gamma$ phase is possible (the possibility of PT outside of Hultgren extrapolation is also discussed [131]). The line $A_{cm}$ intersects also the paraequilibrium line $T_0$, so the different transformation kinetics in the bainitic region above and below $A_{cm}$ can be expected. This may be relevant to the formation of several microstructures below the line $T_0$, such as acicular ferrite and various morphologies of bainite. At last, the line $T_1$ describing the start condition of the $\gamma \to \theta$ transformation lies in the region of high carbon concentration ($c \sim 0.20$), which causes a problem in describing the nucleation of cementite. The possible mechanism of facilitation of cementite formation due to local magnetization and appearance of the intermediate state (MIS) near the boundary of ferrite plate is discussed in Section 7.2.

To conclude this section, it should be stressed that the curves $A_3$, $A_{cm}$, $T_0$, $T_1$, $T_F$ do not depend on the energy relief along the Bain path and are determined only by the terminal values $g_{\gamma[\alpha]}^{PM(FM)}$. On the contrary, the martensitic curves $M_S$ and $M_{S'}$ depend on the transformation energetics at intermediate deformation $e_t$. For the carbon concentration under consideration the magnetic order effects in $\gamma$-Fe are negligible, for the temperatures above $T \sim 400K$. Therefore, the transformations diagram is determined, first of all, by the evolution of magnetic state in $\alpha$-Fe. In particular, the $\gamma \to \alpha$ transition turns out to be possible above Curie temperature ($T_C^\alpha \approx 1043K$) in pure iron due to the short-range ferromagnetic order in $\alpha$-Fe. The short range magnetic order in $\gamma$-Fe becomes important at $T \approx 400K$, which determines the start temperature of the athermal martensitic transformation $M_S$, developing via the lattice instability. Thus, the temperature dependence of magnetic short-range order is the key factor determining the diversity of phase transformations in iron and steel. The closeness of the Curie temperature in $\alpha$-Fe to the temperature of structural transformation is not accidental, but is related with the essence of phase transformations in iron and steel.

# 7. MODELING OF THE PHASE TRANSFORMATION KINETICS

The phase diagram discussed above determines the conditions of the start of phase transformations and the fraction of a new phase at large times. However, it does not allow to understand the intermediate stages of transformation and features of microstructure formation. Namely, the microstructure formed in the intermediate stages of the transformation is a matter of the greatest interest to achieve desirable properties.

## 7.1. Athermal and isothermal martensite transformation.

Phase-field simulation of transformation kinetics in the framework of proposed model, i.e. numerical solution of Eqs. (6), (8) with G.-L. functional (18), is required for the analysis of a martensitic transformation (MT). Herewith, the shear transformation occurs with a velocity ~ $10^3$ m/sec, i.e. in a time much shorter than the characteristic diffusion times. Thus, the carbon distribution can be considered as a "frozen", and the Eq.(8) may be neglected. The modeling of MT kinetics was carried out on a square grid by the classical Runge-Kutta method with periodic boundary conditions [12].

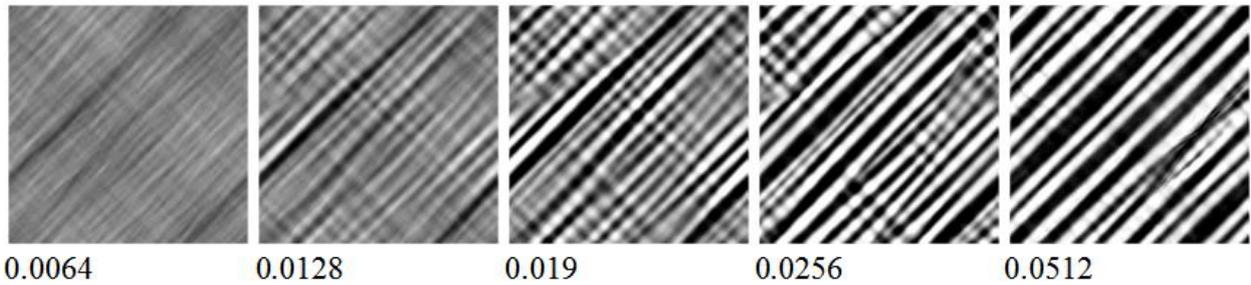

Fig.12. Kinetics of athermal martensitic transformation; $T$=400K, $c$=0. [12]

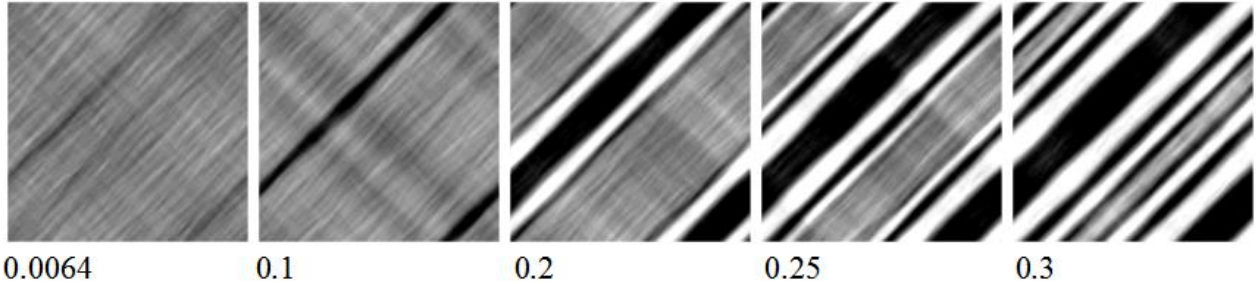

Fig.13. Kinetics of isothermal martensitic transformation; $T$=800K, $c$=0. [12]

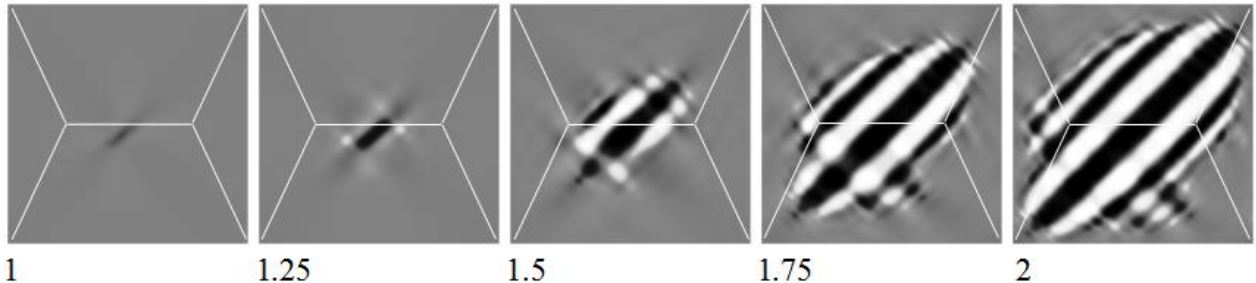

Fig.14. Heterogeneous nucleation of isothermal martensite, with taking into account the relaxation of elastic stresses; $T$=700K, $c$=0.01. [13]

As was mentioned above, the formation of martensite does not require thermal activation at $T<M_S$, while the appearance of isothermal martensite is expected in the temperature range $M_S<T<M_{S'}$ after aging as a result of thermal fluctuations. Therefore, the simulation of MT must take into account the thermal lattice vibrations. The lattice temperature was introduced in the framework of microcanonical ensemble. First, the system is heated to a high temperature ($T$=1200K) by the small random forces $\xi(\mathbf{r},t)$ (the corresponding term is appended to the right side of Eq.(6) leading to the Gibbs distribution of atomic displacements). Then the random forces are shut down and the equilibrium state is attained after aging at this temperature. Finally, the lattice temperature is reduced to the desired value from the interval 400K...1000K by rescaling of velocity field. Herewith, the estimation of lattice temperature was carried out by calculation of the average kinetic energy per degree of freedom, $kT = \rho\Omega <v^2>/2$, where $<v^2>$ is the average square of the velocity over the calculation area. The spin temperature was chosen in the region of stability of the $\gamma$ phase during these preparation procedures and then it switches to the value of lattice temperature.

The typical patterns of the order parameter distribution $\phi$ depending on time are shown in Fig.12–14. Black and white colors correspond to the two possible values of order parameters for the $\alpha$ phase in two-dimensional case, $\phi = \pm1$, i.e. to the two mutually orthogonal directions of the Bain deformation. Time is given in dimensionless units, $t \to t\sqrt{\widetilde{J}_\alpha /(L^2\rho)}$. At significant overcooling ($T<M_S$) the homogeneous transition is realized by development of all fluctuations inherited upon cooling from a high temperature state (see Fig.12). In the temperature range $M_S<T<M_{S2}$ the system remains stable with respect to small fluctuations, and the phase transformation starts with the appearance of critical fluctuation after the incubation period (a few nanoseconds) and it is realized by replication of twinned plates (see Fig.13). The similar mechanism was early observed in Ref.[11] at phase field simulation of MT in the system with an α phase nucleus initially introduced in it. The difference of orientation of the neighboring domains reduces the elastic energy of the system, but raises its surface energy. As a result, the characteristic size of domain is determined by minimization of the sum of elastic and surface contribution to the total energy. The temperature $M_{S2}$ is defined in this case as a result of phase-field simulations in 2D model and it does not necessarily coincide with $M_S$, which was early obtained (see Fig.11a) from the fitting to experimental data. Nevertheless, these temperatures are close; the values of $M_{S2}$ obtained from simulations are designated by the solid circles in Fig.11a.

The elastic stresses play a crucial role in martensite transformation. As it is well known, the components of the deformations tensor are coupled to each other by Saint-Venant's compatibility equations (4). It leads to an additional contribution to the G.-L. functional and results in effective long-range interactions in distribution of order parameters, which play a crucial role in pattern formation upon the phase transition [9–11].

As was shown in [111–114], the relaxation of elastic stresses during the transformation is an important factor determining the martensite morphology. The main relaxation channel is the plastic deformation arising under local stresses exceeding the yield stress. A consequent description of the plastic deformation requires an essential complication of the model by including additional order parameters. Instead, in Ref. [13] the plastic deformation has been taken into account in a phenomenological way. Since the contribution of the elastic stresses to the Ginzburg-Landau free energy functional is determined by the coefficients $A_v$, $A_s$, the real values of these parameters were replaced by some effective, temperature-dependent values, $0<A_v^{\mathrm{eff}}<A_v$, $0<A_s^{\mathrm{eff}}<A_s$. The thermal lattice fluctuations can not be taken into account in this scheme, since the renormalization of $A_v$, $A_s$ leads to the incorrect change of fluctuations amplitude, affecting the start condition of homogeneous transition and the morphology of martensite. This approach can be considered as reasonable for the stage of growth of isothermal martensite. It provides the fast stress relaxation and the lattice remains coherent during the whole transformation process.

Fig.14 shows the MT kinetics when choosing parameters $A_v^{\mathrm{eff}}, A_s^{\mathrm{eff}}$ in such a way that the average elastic energy over the sample is equal to the experimental value of the stored energy in martensite, 0.007eV/at [132] (i.e. ~10% from the nominal values). Herewith, the heterogeneous nucleation is provided by additional contribution to the free energy near grain boundary (see details in Ref.[13]). In this case the martensite is formed as a lenticular colony of twinned plates.

Thus, the proposed model [12,13] reveals two types of MT kinetics, athermal and isothermal at different temperatures, in accordance with existing concepts [133, 2, 22–25]. The experimentally known line of the start of MT [18] corresponds to isothermal MT, whereas the athermal scenario is more hypothetical and it can develop due to short-range magnetic order in the γ phase.

## 7.2. Kinetics of pearlite transformation. Globular and lamellar structures.

The pearlite morphology is similar to that formed during discontinuous decomposition [134–136], when the supersatured mother phase $\alpha_0$ decomposes into a two-phase structure $\alpha_0 \rightarrow \alpha + \beta$, where the phases $\alpha_0$ and $\alpha$ have the same crystal structure, but with different composition. The model of the spinodal decomposition (SD) provoked by grain boundaries was proposed for explanation of this phenomenon [137]. However, the SD kinetics is not applicable to PT, because the mixing energy of carbon in γ-Fe is positive ($v>0$) [57,58] that prevent of the carbon inhomogeneity formation in the γ phase. PT is also similar to the eutectic colony growth in the absence of a temperature gradient [108,109], a process that presupposes a realization of the condition of the solid solution decomposition. As it was shown recently [14], the lamellar structures can arise by an autocatalytic mechanism below the critical temperature even if the mother phase (austenite) is stable with respect to decomposition ($v>0$) and the transition from lamellar structure to globular one takes place with temperature increase.

First of all, let us discuss the possible transformation scenarios of decomposition by using schematic Fig.15 where free energies of involved phases at different temperatures are presented. At a high temperature the stable equilibria of α/γ or γ/θ exist in the system and ferrite or cementite will precipitate from a $\gamma$ matrix in different concentration intervals (curve a). At a lower temperature, $T < T_{eutec}$, the $\gamma$ phase is metastable with respect to the decomposition $\alpha + \gamma$ or $\gamma + \theta$ and, in addition, the stable equilibrium of $\alpha/\theta$ appears (curve b). The uncorrelated nucleation and growth of the $\alpha$ and $\theta$ phases inside a $\gamma$- phase matrix are expected in this case.

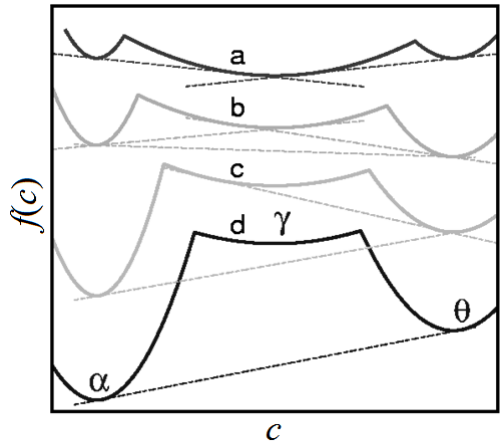

Fig.15. Variants of phase equilibrium in the system with the triple-well potential $f(c)$.

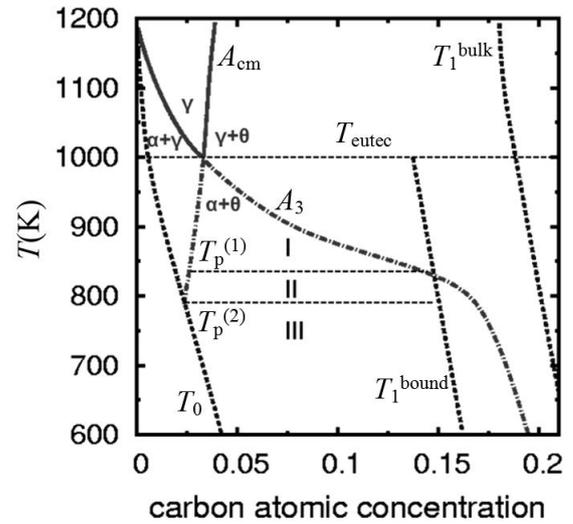

Fig.16. Transformation diagram of carbon steel taken into account the facilitation of cementite formation near the ferrite boundary. The temperature regions I–III are determined by intersection points of the two phase region boundaries $A_3$ and $A_{cm}$ with the paraequilibrium lines $T_0, T_1^{bound}$.

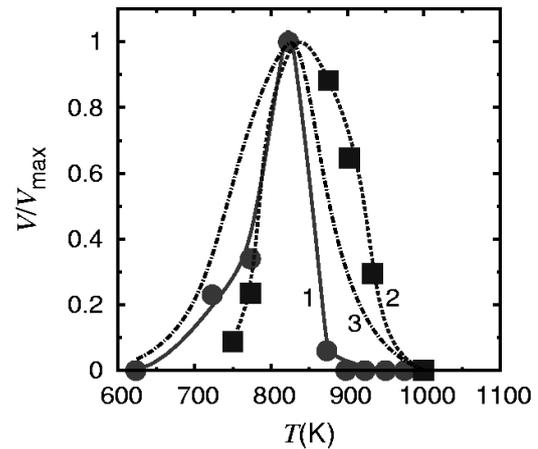

Fig.17. Temperature dependences of pearlite colonies (1) nucleation rate, (2) growth rate, and (3) the effective transformation rate [3].

A further decrease in temperature leads to the loss of one or both metastable equilibria at a preservation of thermodynamic equilibrium between the $\alpha$ and $\theta$ phases (curves c and d). As a result, the change in the decomposition kinetics of austenite is expected, because the cooperative formation of the $\alpha$ and $\theta$ phases becomes preferable. As was shown in Ref. [14], in this case the pearlite colony can emerge by some kind of autocatalytic mechanism when the appearance of one of the phases ($\alpha$ or $\theta$) stimulates the nucleation of next one, and the lamellar or globular structure can form in dependence of the temperature. The similar autocatalytic decomposition scenario was earlier considered for the system with the metastable phase and a symmetric phase diagram, and the possibility of such a mechanism of PT was discussed in Refs [138,139].

A key component of the model PT is the start condition of the cementite formation. Indeed, according to the transformation diagram in Fig.11b, when a carbon concentration increases the metastable equilibrium of $\alpha/\gamma$ is achieved before then the cementite formation is realized, since the line $A_3$ goes much left of $T_1$. To solve this problem, in Ref. [14] it was accepted that cementite nucleation is facilitated in the thin ferromagnetic region near the ferrite plate where so called the Metastable Intermediate Structure (MIS) [77] exists. As a result, the line $T_1$ is shifted to the left by the value $\Delta c_{bound} \sim 0.05$ and crosses the line $A_3$ at approximately 15% carbon. (see Fig.16). Thus, the occurrence of MIS [77], which is a precursor of cementite formation, is very important for the kinetics of PT.

On the transformation diagram in Fig.16 there can be identified three regions I–III where different PT scenarios can be realized in dependence of temperature. These regions are determined by intersection points of the lines $A_3$ and $A_{cm}$ with the lines $T_1^{bound}$ and $T_0$ and correspond to free energies curves b,c,d in the Fig.15, respectively.

The formation of the regular lamellar pearlite by autocatalytic mechanism is to be expected in region III, wherein the instability of austenite in respect to the $\gamma \rightarrow \alpha + \theta$ decomposition appears stepwise with decreasing temperature. This agrees with the experiment [3]; the temperature dependence of pearlite colony nucleation rate, growth rate, and the effective transformation rate (see Fig.17) have a maximum near the temperature 820K, which indicates the thermodynamic instability of austenite. Moreover, the pearlite nucleation rate (in contrast to the growth rate) is close to zero at $T_p^{exp} < T < T_{eutec}$ and changes abruptly at $T_p^{exp} \approx 820K$ (similar results were found earlier in Refs. [53, 66]). So, the nucleation rate appears very slow above 820K, while existing pearlite colony can grow. Therefore, the temperature $T \sim 820K$ may be considered as an experimental estimation of the value $T_p^{(2)}$ in Fig.16.

Since PT is realized above the paraequilibrium temperature ($T > T_0$), it is controlled by carbon diffusion. In this case Eq. (8) can be solved under the assumption that the lattice reconstruction is a rather fast process in comparison with the characteristic diffusion times. In this case, the fast variables $e_t, \eta$ can be avoided by minimization of the local free energy density over these ones, so $f(e_t, \eta, c, T) \rightarrow f_{eff}(c, T)$. In result, the G.-L. functional have a form:

$$F = \int \left( f_{eff}(c, T) + f_{el} + \frac{k_c}{2} (\nabla c)^2 \right) dr \quad (25)$$

$$f_{eff}(c, T) = \min \left\{ f_\alpha(c, T), f_\gamma(c, T), f_\theta(c, T) \right\},$$

where $f_{\gamma(\alpha, \theta)}(c, T)$ is the local density of the free energy of austenite (ferrite, cementite). Since the $\alpha$ and $\theta$ phases in pearlite colonies are usually conjugated with small mismatch and the coherency is lost mostly on the transformation front [140] the elastic energy contribution $f_{el}$ was neglected in [14].

Fig.18 shows the typical evolution of transformation patterns arising at overcooling of austenite into the region III of the transformation diagram. Carbon is pushed out from an embryo of ferrite because its solubility in the $\alpha$ phase is much lower than in the $\gamma$ phase. Since $c(A_3) > c(T_1^{bound})$ (see Fig.16, the region III), the local metastable phase equilibrium of $\alpha/\gamma$ can not be reached, and the formation of cementite takes place. The growth of the arising cementite nucleus leads to depletion of carbon in surrounding austenite. Since $c(A_{cm}) < c(T_0)$ (see Fig.16, the region III), the local metastable phase equilibrium of $\theta/\gamma$ also can not be reached, and the new ferrite layer is formed near the $\theta$ phase. The process described above is repeated, so the corresponding mechanism can be considered as autocatalytic. Phase-field simulation shows that a fine lamellar structure is formed in this case and the movement of the front of pearlite colony is accompanied by increasing its transverse size. As a result, the pearlite colony becomes of a fan-type shape in accordance with experiment [3, 53, 74]. Note that a similar fan-type pearlite structure appears, if we start from one cementite embryo instead of the case of ferrite.

Fig.19 shows the decomposition kinetics in the case, where the metastable equilibrium of the $\gamma$ phase with cementite exists in the region II, but its equilibrium with ferrite is impossible, i.e. $T_p^{(1)} > T_p^{(2)}$. In this case the PT starts only with ferrite embryos, since they alone can not be in equilibrium with austenite. The condition of autocatalytic multiplication of lamellae is violated and the phase-field simulation demonstrates the relevance of globular structure. As in the previous case, carbon is pushed out from the embryo of ferrite and the nucleation of cementite takes place. However, in this case the line $A_{cm}$ is achieved before the critical concentration $c(T_0)$ does, so that the metastable phase equilibrium of $\gamma/\theta$ is realized, and the new ferritic layer does not appear. As a result, the other scenario of transformation takes place, which results in numerous small cementite precipitates in the single ferritic matrix.

In the region I in Fig.16 austenite is decomposed by conventional nucleation-and-growth mechanism, as was discussed in Ref.[13], and we do not show corresponding pictures here. Carbon is pushed out from the ferrite embryo and its concentration near ferrite interface reaches the value determined by $A_3$ curve. Since $c(A_3) < c(T_1^{bound})$, the metastable phase equilibrium of $\alpha/\gamma$ is reached, and the formation of cementite does not occur in this case. And vice versa, if we start from one cementite embryo, the metastable phase equilibrium of $\gamma/\theta$ is realized and ferrite does not occur because $c(A_{cm}) > c(T_0)$.

Thus, the two possible scenarios of pearlite transformation, lamellar and globular, are possible within the model presented in [14], and second one is realized at a higher temperature. The autocatalytic decomposition described above differs from the well-known spinodal decomposition (SD) by the fact that the $\gamma$ phase loses its stability in respect to large composition deviations (near the existing precipitates), so that decomposition is realized by the scenario of colonies growth, while during SD the homogeneous instability of solid solution in respect to small compositional fluctuations develops in the bulk.

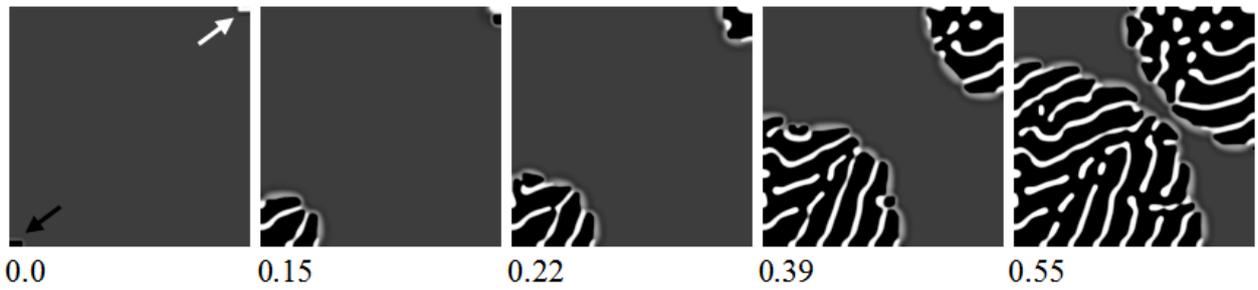

| 0.0 | 0.15 | 0.22 | 0.39 | 0.55 |

Fig.18. Kinetics of lamellar structure growth from a nucleus placed on the grain boundaries junctions (ferrite nucleus on the bottom left and cementite nucleus on the upper right are indicated by arrows); $T$=675K, $c_0$=0.06, $v_\gamma$ =1.5 eV/at [14]. The carbon concentration is indicated by the gray scale; the black color corresponds to ferrite, and the white to cementite. The time is given in dimensionless units, $L^2/D_\alpha$. The embryos of ferrite and cementite are introduced into the initial state, lower left and upper right corner of the calculation square, respectively.

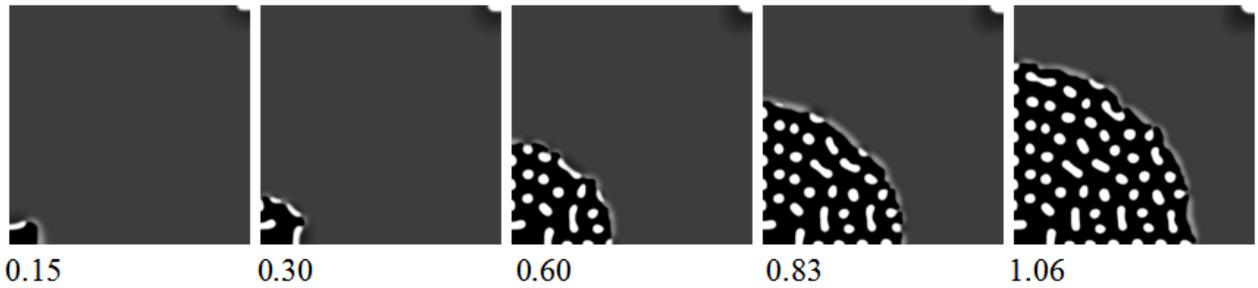

| 0.15 | 0.30 | 0.60 | 0.83 | 1.06 |

Fig.19. Kinetics of globular colony growth from a ferrite nucleus; $T$=800K, $c_0$=0.06, $v_\gamma$ =1.5 eV/at [14].

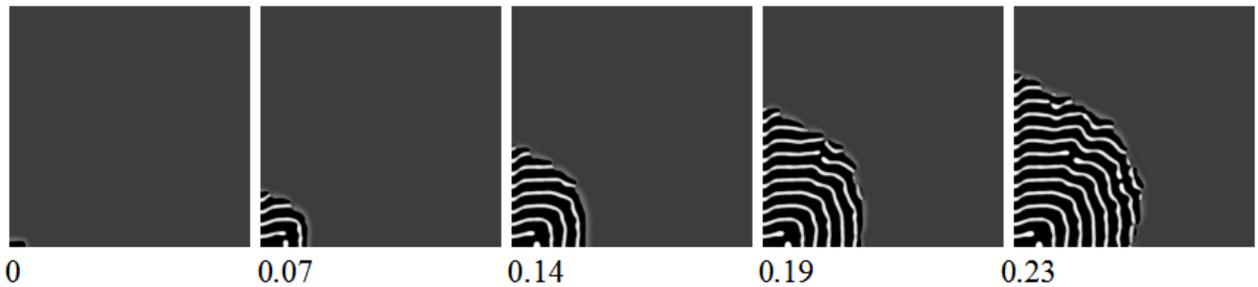

| 0 | 0.07 | 0.14 | 0.19 | 0.23 |

Fig.20. Kinetics of lamellar structure growth from a nucleus of ferrite; $T$=675K, $c_0$=0.06, $v_\gamma$ =2 eV/at.

The nucleation of globular pearlite, also known as Divorced Eutectoid Transformation (DET), attracts an essential interest [71–75]. This state is usually produced by the heating of the existing lamellar pearlite above the temperature $T_{eutec}$ until the cementite is almost completely dissolved, and then the cooling below the temperature $T_{eutec}$ is carried out. As a result, the observed PT morphologies is similar to the Fig.19, wherein the numerous precipitations of cementite are immersed in the single $\alpha$ matrix with a pronounced transformation front. According the conventional point of view, the cementite nucleuses are storing in the $\gamma$ matrix after the heating and grow upon a subsequent small overcooling below $T_{eutec}$, while the nucleation of lamellar structure do not occur before completion of DET. This scenario is consistent with the transformation in the region I (see Fig.16). Moreover, in Ref.[72] it was pointed out that the globular pearlite is realized in hypoeutectoid steels even at overcooling from an almost homogeneous state, thus the number of cementite globules after the DET is much more than the number of potential nuclei. In the context of presented phase-field simulations (Fig.19), this fact may indicate that the kinetics of globular pearlite includes the autocatalytic nucleation of the new cementite globules, as it occurs in the region II of transformation diagram.

Variation of the parameters leads to some changes of the precipitates morphology. We only discuss the general trends observed in the calculations. The interlamellar spacing decreases with the decreasing of temperature $T$ in accordance with known classical concepts [3]. The ratio of the temperatures $T_p^{(1)}, T_p^{(2)}$ can be changed by varying of the parameters $\varepsilon_\gamma^{FM(PM)}$ and $v_\gamma$. The tendency of lamellar structure formation increases with increasing $v_\gamma$ (see Fig.20), however, the morphology of lamellae differs from the conventional pearlitic structure (the concentric layers instead of radial strips are observed). In our opinion, the elastic stresses can play an essential role in the orientation of lamellae, which are not taken into account in the simplified G.-L. functional, Eq.(25).

The qualitative conclusions presented here are quite general and they can be attributed to other eutectoid systems, for example to the alloy Zn-Al [50], where the lamellar structures are also formed. In the same time, the proposed model does not explain the appearance of a small number of colonies of coarse lamellar pearlite, which is observed in the temperature range $T_p^{exp} < T < T_{eutec}$ [3], i.e. together with DET. So, additional factors should be taken into account (such as incompatibility elastic stresses) to provide more reliable results of calculations.

### 7.3. Scenarios of ferrite and intermediate transformations.

The ferrite transformation (FT) starts just after the cooling below the line $A_3$ and results in the appearance of almost pure bcc-iron ($\alpha$ phase). Because driving force in this case is rather small, the transformation usually starts on grain boundaries where nucleation is facilitated. Since carbon solubility in the $\alpha$ phase is very small, the carbon is pushed out into the $\gamma$ matrix, which results in the appearance of the regions depleted or enriched in carbon. FT is a diffusion-controlled phase transformation, so that nucleus of the $\alpha$ phase can not grow without carbon redistribution in this case. The temperature region of diffusion controlled growth of the $\alpha$ phase is $T_0 < T < A_3$ (see Fig.11b) and the condition where ferrite can grow without cementite formation is $T > T_{eutec}$.

It is necessary to pay attention to the two important features of FT. At first, the gain of free energy in the formation of ferrite is small (see Fig.9b), therefore the realization of FT requires almost complete relaxation of elastic stresses. Secondly, FT is even observed experimentally above the Curie temperature, $T > T_C$, thus it is due to short-range magnetic order in the absence of long-range order.

Fig.21 shows the kinetics of FT when the solution of the complete set of equations for shear-diffusion transformations is carried out. The upper and lower rows of images correspond to the shear order parameter and carbon concentration, respectively. Time is given in dimensionless units, $t \to t\sqrt{\tilde{J}_\alpha}/(L^2\rho)$. It was supposed, the elastic stresses are absent, $A_v^{eff} = A_s^{eff} = 0$, and the additional contribution to the free energy exist near the grain triple junctions and boundaries (see [13] for details). The growing polygonal ferrite precipitates, surrounded by a carbon shell, are observed in accordance with experiments [6].

In the temperature range $M_{S'} < T < T_0$ the model demonstrates several possible scenarios. As was noted in the discussion of a transformation diagram, at temperature $T < T_0$ the lattice reconstruction $\gamma \to \alpha$ can occur even if homogeneous carbon distribution is retained. However, the diagram (Fig.11) was constructed without taking into account the contribution of elastic stresses to the free energy. As was discussed in [13], the nominal contribution of elastic stresses is very large, but it decreases due to plastic deformation. The effective values $A_v^{eff}, A_s^{eff}$ can be roughly estimated from experimental data on the residual stresses [4]. As a result, the start temperature of shear transformation decreases, $T_{start} < T_0$. Therefore, transformation scenario in the temperature range $M_{S'} < T < T_0$ depends on degree of undercooling. Here we restrict ourselves to discussion of the scenarios of ferrite nucleation and growth in the presence of elastic stresses, without the cementite formation, and we call them scenarios of intermediate transformation.

At elevated temperatures the transformation is controlled by carbon diffusion, as in the case of ferrite formation, but the precipitate of $\alpha$ phase acquires a plate form, like bainitic or Widmanstatten ferrite (Fig.22). At lower temperatures the ferrite nucleus appears and grows by shear mechanism, up to some critical size, determined by increasing elastic stresses (Fig.23). In this case the depletion of carbon of an $\alpha$ plate and its diffusive growth occurs in the second stage. Besides, at the same temperatures, but at a large amplitude of the initial perturbation on GB, the sheaf of ferrite plates can be formed, in which the elastic stresses produced by tetragonal deformations are compensated (Fig.24). For large times, the carbon accumulates at the boundaries of ferrite plates, where its concentration reaches the big values, and cementite nuclei may appear.

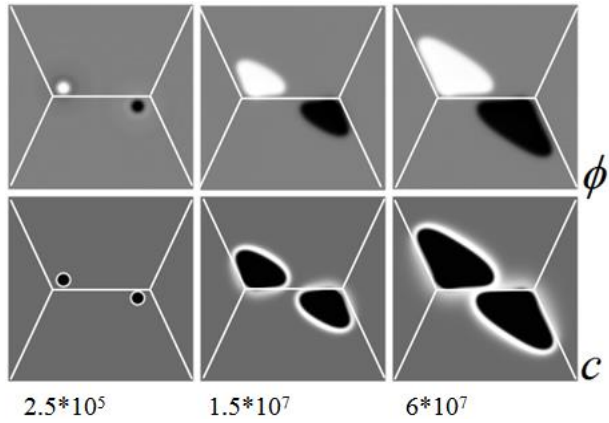

Fig.21. Kinetics of formation of polygonal ferrite at triple grain junctions, $T$=1000K, $c_0$=0.01 [13].

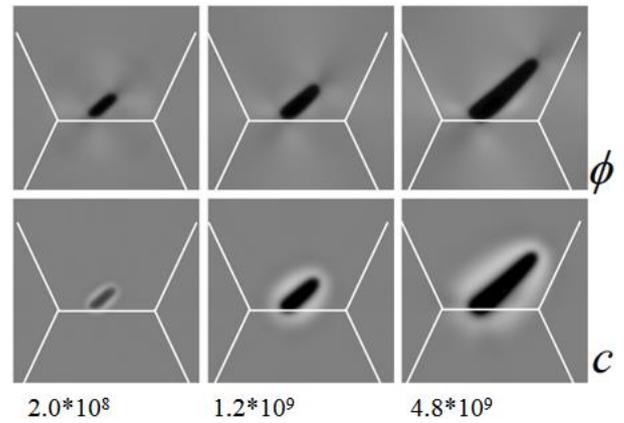

Fig.22. Diffusion-controlled nucleation and growth of bainitic ferrite plate with taking into account the elastic stresses, $T$=850K, $c_0$=0.01.

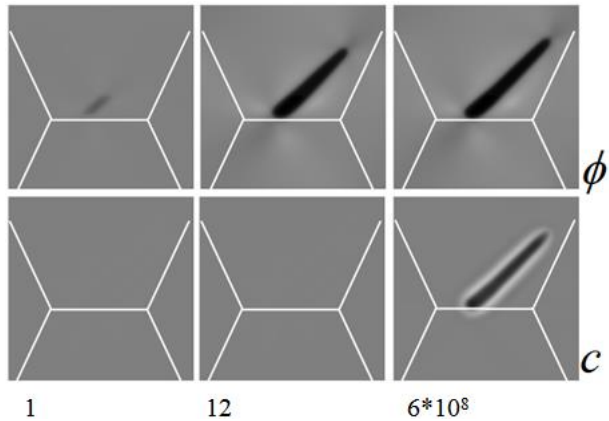

Fig.23. Shear-controlled nucleation and growth of bainitic ferrite plate with taking into account the elastic stresses, $T$=800K, $c_0$=0.01.

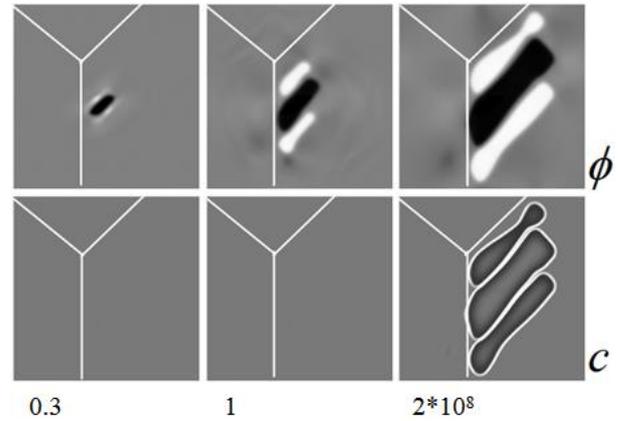

Fig.24. Shear-controlled nucleation and growth of the sheaf of bainitic ferrite plates, $T$=800K, $c_0$=0.01[13].

The scenarios of intermediate transformation shown in Fig.22–24 may be associated with upper and lower bainite. Indeed, the opinion that upper bainite emerges by diffusion and lower bainite by shear mechanism is widespread [19]. However, the stopping of growth of plates in the presented calculation is due to the boundary conditions, whereas the stopping of bainitic plate growth is explained by lattice coherency disturbance on the boundary γ/α after the plastic deformation [4]. Herewith, all subunits in the bainite colony have the same orientation; this is problematic without coherency disturbance on the colony front.

Thus, the proposed model describes the main peculiarities of ferrite and some features of initial stages of bainite transformations. The consistent model of bainite transformation should take into account the plastic deformation more correctly, including the loss of lattice coherency at the γ/α boundary when the critical size of ferrite subunits is reached.

## 8. EFFECT OF EXTERNAL MAGNETIC FIELD ON THE START OF PHASE TRANSFORMATIONS

The effect of powerful pulsed magnetic field on the martensitic transformation (MT) in steel was first discovered in Ref.[141]. In Ref. [142, 143] it was shown that the magnetic field linearly shifts the start temperature of MT ($M_S^{exp}$ increases by about 0.5 degree in the field $H$=1kOe). In Ref. [144, 145] it was concluded that the pulsed magnetic fields do not affect the rate of athermal MT, but can provoke the athermal MT, leading to the specific distinctive morphology of martensitic crystals. This was explained as follows: the rate of athermal MT is close to the impulse duration (~$10^{-3}$ sec), while the rate of isothermal MT is much less (ten minutes), so that isothermal MT can be realized only in a powerful static field; such fields were not available in Ref. [144, 145]. In Ref. [146] it was shown that the static field shifts the start temperature of isothermal MT. Futher investigation has pointed out that the static magnetic field 50kOe also accelerates the pearlite and bainite transformations, herewith the start temperature shifts by 10 degrees [145]. The interest to the effect of external magnetic field on the kinetics of diffusion controlled transformations increases in recent years [147–150]. In particular, it was found that the magnetic field enhance the mass fraction of proeutectoid ferrite and influences the morphology of cementite precipitates.

To explain the effect of magnetic field on MT, the Krivoglaz-Sadovsky equation was proposed [145, 151]. According to this formula, the magnetic field shifts the thermodynamic equilibrium to the formation of a magnetic α phase

$$\Delta T = T_0 V_\alpha M_\alpha H / q, \tag{26}$$

where $T_0$ is the start temperature of the $\gamma \rightarrow \alpha$ transformation, $V_\alpha$, $M_\alpha$ are the volume and the magnetization of the α phase, $H$ is the magnetic field, $q$ is the heat of transformation. Also, the formula for the shifting of solubility limits was obtained from the condition of equality of chemical potentials of the phases:

$$\frac{c_1(H)}{c_2(H)} = \frac{c_1(0)}{c_2(0)} \mathrm{Exp}\left(\frac{H}{RT}\frac{\partial(M_1 V_1)}{\partial c}\right) \tag{27}$$

where $c_i(H)$, $c_i(0)$ are solubility limits in the magnetic field and without it.

It should be noted that the Eqs (26), (27) correspond to the lines $T_0$, $A_3$ (see Fig.11) determined by equilibrium conditions, while the lines $M_S$, $M_{S'}$ are related to the barrier on the Bain path. Thus, formulas (26), (27) can not be used for the analysis of a martensitic transformation, contrary to the popular belief. According to the model [13], the athermal MT is due to the appearance of short-range magnetic order in the $\gamma$ phase, and isothermal MT depends on the energy of some intermediate state near the $\gamma$ phase on the Bain path.

The transformation diagram in the case of the presence of external magnetic field is presented in Fig.25. Here we used the general formulas (10)–(11) with an additional contribution $-g\beta H_0\sigma$ and formulas (13) –(15) for the spin correlation function. The external magnetic field increases the degree of order near the Curie temperature, which results in increase of the spin correlator magnitude and the shift of lines of the transformation diagram. Herewith, the change of the magnitude of the correlator depends on a tetragonal deformation,

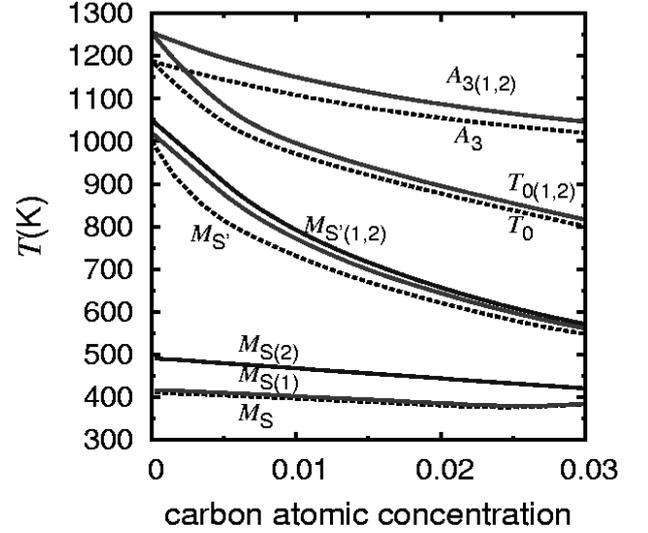

Fig.25. The transformation diagram "under" external magnetic field, $H$=50kOe. The lines $A_{3(1)}$, $T_{0(1)}$, $M_{S(1)}$, $M_{S'(1)}$ and $A_{3(2)}$, $T_{0(2)}$, $M_{S(2)}$, $M_{S'(2)}$ correspond to the choice of $f_s(\phi) = (1 - \phi^2)^2$ or $f_s(\phi) = 1 - |\phi|$, respectively. The dotted lines $A_3$, $T_0$, $M_S$, $M_{S'}$ correspond to the absence of external field.

$$\Delta Q^H(T, \phi) = \Delta Q_\gamma^H + (\Delta Q_\alpha^H - \Delta Q_\gamma^H)(1 - f_s(\phi)), \tag{28}$$

where $\Delta Q_{\gamma(\alpha)}^H$ is the change of correlator in $\gamma(\alpha)$ phases, and $f_s(\phi)$ is the function characterizing the magnetic susceptibility in intermediate lattice states. Fig.26 shows the transformation diagram "under" the field $H$=50kOe in the case of $f_s(\phi) = (1 - \phi^2)^2$ (curves $A_{3(1)}$, $T_{0(1)}$, $M_{S(1)}$, $M_{S'(1)}$) and $f_s(\phi) = 1 - |\phi|$ (curves $A_{3(2)}$, $T_{0(2)}$, $M_{S(2)}$, $M_{S'(2)}$). One can see that the lines $A_3$, $T_0$ do not depend on the choice of $f_s(\phi)$, whereas the lines $M_S$, $M_{S'}$ are very sensitive to this.

Thus, the proposed model allow us to find the shifts of the lines $A_3$, $T_0$ in agreement with Refs.[145], and it also shows that the formulas (26), (27) can not be used for the lines $M_S$, $M_{S'}$. The construction of these lines under external magnetic field is to be a separate problem and requires a justification of the form of $f_s(\phi)$.

## 9. CONCLUSIONS AND OUTLOOK

The problem of the phase transformations and microstructure formation in iron and steel is to be in scope of interest for a long time and is actively discussed by now [1–6, 19, 22]. Nevertheless, despite great efforts, the number of important questions are still under debates. One of the reasons for this is the complexity of the phase transformations in iron based alloys that involve both lattice and magnetic degrees of freedom, as well the carbon redistribution, which also plays an important role. Besides, the processes of transformation involve several spatial scale levels, from microscopic (atomistic) to macroscopic (at the level of the grain size).

Starting with a conceptual work by Zener [37], it is believed that magnetism plays a crucial role in the phase transformations in iron and steels. However, all early proposed models are too phenomenological, so their correct choice is impossible. In this review we have presented recent progress in understanding of microscopic mechanisms of phase transformations in iron and steel. This progress was possible, on the one hand, due to the widely using of ab initio methods for calculation of the electronic structure and total energy in different structural and magnetic states of iron [7, 8, 35, 36, 58, 77, 120, 126, 152], and, on the other hand, due to applications of the atomistic simulations within the phase-field approach [81] to the transformations kinetics [9–14, 48, 74, 75, 111, 112].

The rapid development of computing technology offers the prospect for the research of realistic transformations kinetics in 3D-models depending on cooling rate and concentrations of alloying elements. In that connection the task of constructing of the consistent ab initio based model of phase transformations in steel, describing the shear-diffusion transformation kinetics and taking into account the magnetic degree of freedom is very relevant.

The recently proposed model [12–14] agrees well with the known experimental data and predicts the start temperatures of different transformations (ferrite, pearlite, bainite, martensite). It was shown that the magnetism provides the main contribution to the change of free energy at the $\gamma \rightarrow \alpha$ transformation. Therefore, the increase of short-range magnetic order plays a key role in the change of transformation scenarios (from ferrite to martensite) under cooling. Phase-field simulation carried out in the framework of the proposed model reproduces the typical precipitates morphology, including ferrite, twinned martensite, and pearlite colonies.

The ferrite transformation starts at a temperature below $A_3$ due to the short-range magnetic order (with a possible absence of long-range order) and requires the essential relaxation of the elastic stresses. The pearlite transformation results in the formation of a regular structure due to autocatalytic mechanism, which is realized in the absence of thermodynamic equilibrium between initial austenite and transformation products (ferrite and cementite). Two types of autocatalysis were revealed leading to a lamellar or globular pearlite structure depending on the temperature. Also two types of intermediate

(bainite) transformations were observed below the paraequilibrium temperature $T_0$ in phase-field modeling. These are diffusion and shear-controlled transformations, which can be associated with upper and lower bainite, respectively. The experimental curve of the start of martensitic transformation (MT) corresponds to the conception of isothermal martensite, whereas the classical (athermal) scenario of MT is due to the short-range magnetic order in $\gamma$-Fe, which arises at lower temperature. The model allows us to consider an effect of external magnetic field on the curves of the start of ferritic and bainitic transformations in agreement with the Krivoglaz-Sadovsky concept [145], and reveals inapplicability of this concept to a martensitic transformation.

Despite the significant progress in recent years, a number of problems remain unresolved, including peculiarities of bainitic microstructure with taking into account the cementite formation, the role of elastic stresses, and their plastic relaxation in growth kinetics of the pearlite and bainite colonies, the effect of alloying elements on the thermodynamics and kinetics of phase transformations. Also the dimension of model (2D or 3D) is essential for the kinetics [28, 29, 87, 112], so that more realistic simulations should be based on 3D models.

According to modern views [4] the role of plastic deformation increases with temperature; this is a principal channel of elastic energy relaxation in the case of ferrite transformation, while the relaxation of elastic energy in the case of martensitic transformation is provided by twinning of plates. Herewith, plastic deformation causes the bainite morphology, since the characteristic size of the bainitic subunits is determined by the start condition of plastic deformation, disturbing the lattice coherency at the interface $\gamma/\alpha$.

Thus, the essential contours of an ab initio based theory of phase transformations in iron and steel are formed. The further development of the theory and its applications to complex alloyed steels at various cooling regimes should lead to practical applications that are significant for metallurgical production.

## REFERENCES

1. *Kwon O.* What's new in steel? // Nature Materials. 2007. V.6. P.713
2. *Kurdjumov G.V., Utevski L.M., Entin R.I.* Transformation in iron and steel. Moscow: Nauka, 1977. 239 pp. [In Russian]

3. *Schastlivtsev V.M., Mirzaev D.A., Yakovleva I.L.* Pearlite in Carbon Steels. Ekaterinburg: Ural. Otd. Ross. Akad. Nauk, 2006. 311 pp. [In Russian]

4. *Bhadeshia H.K.D.H.* Bainite in steels. London: IOM Communications Ltd, 2001. 460 pp.

5. *Leslie W.C., Hornbogen E.* Physical metallurgy of steels, in Physical Metallurgy. V. 2. Ed. by Cahn R.W., Haasen P. Elsevier, 1996. P.1555–1620

6. *Bhadeshia H.K.D.H., Honeycombe R.W.K.* Steels: Microstructure and Properties. Ed.3. Oxford: Butterworth–Heinemann, 1995. 360 pp.

7. *Okatov S.V., Kuznetsov A.R., Gornostyrev Yu.N., Urtsev V.N., Katsnelson M.I.* Effect of magnetic state on the $\gamma$–$\alpha$ transition in iron: First-principles calculations of the Bain transformation path. // Phys. Rev. B. 2009. V.79. 094111 (4 pp)

8. *Okatov S.V., Gornostyrev Yu.N., Lichtenstein A.I., Katsnelson M.I.* Magnetoelastic coupling in $\gamma$-iron // Phys. Rev. B. 2011. V.84. 214422 (8 pp)

9. *Rasmussen K.Ø., Lookman T., Saxena A., Bishop A.R., Albers R.C., Shenoy S.R.* Three-Dimensional Elastic Compatibility and Varieties of Twins in Martensites // Phys. Rev. Lett. 2001. V.87. №5. 055704 (4 pp)

10. *Bouville M., Ahluwalia R.* Interplay between Diffusive and Displacive Phase Transformations: Time-Temperature-Transformation Diagrams and Microstructures // Phys. Rev. Lett. 2006. V.97. 055701 (4 pp)

11. *Shenoy S.R., Lookman T., Saxena A., Bishop A.R.* Martensitic textures: Multiscale consequences of elastic compatibility. // Phys. Rev. B. 1999. V.60. №18. R12. P.537–541

12. *Razumov I.K., Gornostyrev Yu.N., Katsnelson M.I.* Effect of magnetism on kinetics of $\gamma$–$\alpha$ transformation and pattern formation in iron. // J. of Physics: Cond. Mat. 2013. V.25. №13. 135401

13. *Razumov I.K., Boukhvalov D.V., Petrik M.V., Urtsev V.N., Shmakov A.V., Katsnelson M.I., Gornostyrev Yu.N.* Role of magnetic degrees of freedom in a scenario of phase transformations in steel // Phys. Rev. B. 2014. V.90. 094101 (8 pp)

14. *Razumov I.K., Gornostyrev Yu.N., Katsnelson M.I.* Autocatalytic mechanism of pearlite transformation. Submitted in Phys. Rev. Applied.

15. *Bernshtein M.L., Kurdjumov G.V., Mes'kin V.S., Popov A.A, Sadovsky V.D., Skakov Yu.A., Schastlivtsev V.M., Taran Yu.N., Utevsky L.M.,Entin R.I.* The Iron-Carbon. /In.: Metallurgy and Heat Treatment of steel and cast iron, V.3, Eds. *Rakhshtadt A.G., Kaputkina L.M., Prokoshkin S.D., Supov A.V.* Intermet Engineering, Moscow, 2005 [In Russian]

16. *Okamoto H.* The C-Fe (Carbon-Iron) System. // J. of Phase Equilibria. 1992. V.13. N5. P.543–565 Eisenhüttenw. 1961. V.32. P.251–260

17. *Kaufman L., Radcliffe S.V., Cohen M.* //In: Decomposition of Austenite by Diffusional Processes. Ed. by Zackay V.F. and Aaronson H.I. AIME, New York: Interscience Publishers, 1962.

18. *Liu C., Zao Z., Northwood D.O., Liu Y.* A new empirical formula for the calculation of $M_S$

temperatures in pure iron and super-low carbon alloy steels // J. Mater. Process. Technol. 2001, V.113, P.556–562

19. *Fielding L.C.D.* The Bainite Controversy. // Mat.Sci. and Technology. 2013. V.29. №4. P.383–399

20. *Bhadeshia H.K.D.H., Svensson L.-E.* Modelling the Evolution of Microstructure in Steel Weld Metal. /in: Mathematical Modelling of Weld Phenomena, eds. Cerjak H., Eastering K.E. London: Institute of Materials, 1993. P.109–182

21. *Zener C.* Kinetics of Decomposition of an Austenite // Trans. AIME. 1946. V.167. P.550–595

22. *Kurdjumov G.V.* Non-diffusional (martensitic) transitions in alloys // Doklady AN SSSR 1948. V.60. №9. P.1543–1546

23. *Cohen M., Machlin E.S., Paranjpe V.G.* Thermodynamics of the Martensitic Transformation // in: Thermodynamics in Physical Metallurgy. Am. Soc. Metals, Cleveland, 1950. 242 pp.

24. *Shih C.H., Averbach B.H., Cohen M.* Some Characteristics of the Isothermal Martensitic Transformation // Trans. AIME. 1955. V.203. P.183

25. *Cech R.E., Turnbull D.J.* Heterogeneous nucleation of the martensite transformation // Trans. AIME. 1956. V.206. P.124–132

26. *Bein E.C.* The nature of martensite // Trans. AIMME. 1924. V.70. P.25–46

27. *Kurdjumov G.V., Sachs G.* Over the mechanisms of steel hardering // Z. Phys. 1930. V.64. P.325–343

28. *Barsch G.R., Krumhansl J.A.* Twin Boundaries in Ferroelastic Media without Interface Dislocations // Phys. Rev. Lett. 1984. V.53. №11. 1069–1072.

29. *Krumhansl J.A., Gooding R.J.* Structural phase transitions with little phonon softering and first-order character // Phys. Rev. B. 1989. V.39. №5. P.3047–3056

30. *Hume-Rothery W.* Properties and Conditions of Formation of Intermetallic Compounds // J. Inst. Met. 1926. V.35. P.295–361

31. *Katsnelson M.I., Naumov I., Trefilov A.V.* Singularities of the electronic structure and premartensitic anomalies of lattice properties in beta-phases of metals and alloys // Phase Transitions. 1994. V.49. P.143–191

32. *De Fontaine D, Kikuchi R.* Bragg-Williams and Other Models of the Omega Phase Transformation // Acta Metall. 1974. V.22. P.1139–1146

33. *Cook H.E.* On First-Order Structural Phase Transitions // Acta Metall. 1975. V.23. P.1027–1054

34. *Neuhaus J., Petry W., Krimmel A.* Phonon softening and martensitic transformation in $\alpha$-Fe // Physica B. 1997. V.234-236. P.897–899

35. *Leonov I., Poteryaev A.I., Anisimov V.I., Vollhardt D.* Calculated phonon spectra of paramagnetic iron at the $\alpha$-$\gamma$ phase transition // Phys. Rev. B. 2012. V.85. 020401 (4 pp)

36. *Körmann F., Dick A., Grabovski B., Hickel T., Neugebauer J.* Atomic forces at finite magnetic temperatures: Phonons in paramagnetic iron // Phys. Rev. B. 2012. V.85. 125104 (5 pp)

37. *Zener C.* Elasticity and Anelasticity of Metals. Chicago: University of Chicago Press, 1948

38. *Kaufman L., Clougherty E.V., Weiss R.J.* Lattice stability of metals. 3. Iron. // Acta Metall. 1963. V.11. P.323–335

39. *Hasegawa H., Pettifor D.G.* Microscopic Theory of the Temperature - Pressure Phase Diagram of Iron // Phys. Rev. Lett. 1983. V.50. P.130–133

40. *Boukhvalov D.W., Gornostyrev Yu.N., Katsnelson M.I., Lichtenstein A.I.* Magnetism and Local Distortions near Carbon Impurity in γ-Iron // Phys. Rev. Lett. 2007. V.99. 247205 (4 pp)

41. *Hultrgen A.* Isothermal transformation of austenite. Trans. ASM. 1947. V.39. P.915–1005

42. *Hillert M.* Paraequilibrium. Technical report, Swedish Institute for Metals Research, Stockholm, Sweden, 1953

43. *Klier E.P., Lyman T.* The bainite reaction in hypoeutectoid steels // Trans. AIMME. Met. Technol. 1944. P. 395–422

44. *Ko T., Cottrell S.A.* The formation of bainite // J. Iron. Steel Inst. 1952. V.172. P.307–313

45. *Hillert M.* The growth of ferrite, bainite and martensite. Internal report, Swedish Institute for Metals Research, Stockholm, Sweden, 1960

46. *Hultgren J.* // J. Iron. Steel Inst. 1926. V.114. P.421–422

47. *Christian J.W.* The origins of surface relief effects in phase transformations. //In: Decomposition of Austenite by Diffusional Processes. Ed. by Zackay V.F. and Aaronson H.I. AIME, New York, Interscience, 1962. P.371–386

48. *Hillert M., Hoglund L., Agren J.* Role of carbon and alloying elements in the formation of bainitic ferrite. // Metall. Mater. Trans. A. 2004. V.35. P.3693–3700

49. *Aaronson H.I.* The Mechanism of Phase Transformations in Crystalline Solids. London: The Institute of Metals, 1969. 270 pp.

50. *Ling F.-W., Laughlin D.E.* The Kinetics of Transformation in Zn-Al Superplastic Alloys // Met.Trans. A. 1979. V.10A. P.921–928

51. *Adorno A.T., Benedetti A.V., Da Silva R.A.G., Blanco M.* Influence of the Al content on the phse transformation in Cu-Al-Ag alloys. // Ecletica Quimica. 2003. V.28. №1. P.33–38

52. *Das A., Gust W., Mittemeijer E.J.* Eutectoid transformation in Au-39 at.%In. // J. Mat.Sci and Tech. 2000. V.16. P.593–598

53. *Abbaschian R., Abbaschian L., Reed-Hill R.* Physical Metallurgy Principles. SI Version. Stamford, CT: Cengage Learning, 2009. 750 pp.

54. *Kral M.V., Mangan M.A., Spanos G.* Three-dimensional analysis of microstructures. // Materials Characterisation. 2000. V.45. P.17–23

55. *Graef M.D., Kral M.V., Hillert M.* A modern 3D view of an old perlite colony. // J. Metals. 2006. V.58. P.25–28

56. *Cahn J.W., Hilliard J.E.* Free energy of a nonuniform system. I. Interfacial free energy // J.Chem.Phys. 1958. V.28. P.258–267

57. *Bhadeshia H.K.D.H.* Carbon–Carbon Interactions in Iron. // J. Mat. Sci. 2004. V.39. P.3949–3955

58. *Ponomareva A.V., Gornostyrev Yu.N., Abrikosov I.A.* Energy of interaction between carbon impurities in paramagnetic γ - iron // JETP. 2015. V.120. №4. P.716–724

59. *Hillert M.* Solid State Phase Transformation // Jemkontorets Annaler. 1957. V.141. №11. P.757–790

60. *Turnbull D.* Theory of cellular precipitation. // Acta Metall. 1955. V.3. №1. P.55–63

61. *Sundquist B.E.* The edgewise growth of pearlite // Acta Metall. 1968. V.16. №12. P.1413–1422

62. *Vaks V.G., Stroev A.Yu.* Kinetics of the eutectoid colony growth in a solid solution for simple alloy models // JETP. 2008. V.107. №1. P.90–101

63. *Vaks V.G., Stroev A.Y., Urtsev V.N., Shmakov A.V.* Experimental and theoretical study of the formation and growth of pearlite colonies in eutectoid steels // JEPT. 2011. V.112. №6. P.961–978

64. *Yamanaka A., Yamamoto T., Takaki T., Tomita Y.* Multi-Phase-Field Study for Pearlite Transformation with Grain Boundary Diffusion. IV International Conference Multiscale Materials Modeling (MMM2008) (October 27-31, 2008, Florida, USA)

65. *Ankit K., Choudhury A., Qin C., Schulz S., McDaniel M., Nestler B.* Theoretical and numerical study of lamellar eutectoid growth influenced by volume diffusion. // Acta Mater. 2003. V.61. P.4245–4253

66. *Mehl R.F., Dubé A.* The eutectoid reaction. / in: Phase Transformation in Solids. Ed. by Mayer J.E. Smoluchowski R. and Weyl W.A. New York: John Wiley and Sons, Inc., 1951. P. 545–582

67. *Smith G.V., Mehl R.F.* Lattice relationships in decomposition of austenite to pearlite, bainite and martensite. // Trans. AIME. 1942. V.150. P.211–226

68. *Nicholson M.E.* On the nucleation of pearlite. // Journal of Metals. 1954. V.6. P.1071–1074

69. *Tu K.N., Turnbull D.* Morphology and structure of tin lamellae formed by cellular precipitation // Acta Metall. 1969. V.17. P.1263–1279

70. *Hillert M.* The formation of pearlite / in: Decomposition of Austenite by Diffusional Processes. Ed. by Zackay V.F. and Aaronson H.I. New York: Interscience, 1962. P.197–237

71. *Pandit A.S., Bhadeshia H.K.D.H.* Divorced pearlite in Steels. // Proceedings of the Royal Society A. 2012. V.468. №2145. P.2767–2778

72. *Verhoeven J.D., Gibson E.D.* The divorced eutectoid transformation in steel. // Metallurgical and Materials Transactions A. 1998. V.29. №4. P.1181–1189

73. *Oyama T., Sherby O.D., Wadworth J., Walser B.* Application of the divorced eutectoid transformation to the development of fine-grained, spheroidized structures in ultrahigh carbon steels. // Scripta Metall. 1984. V.18. P.799–804

74. *Ankit K., Mukherjee R., Mittnacht T., Nestler B.* Deviations from cooperative growth mode during eutectoid transformation: insights from phase field approach. // Acta Mater. 2014. V.81. P.204–209

75. *Ankit K., Mukherjee R., Nestler B.* Deviations from cooperative growth mode during eutectoid transformation: Mechanisms of polycrystalline eutectoid evolution in Fe-C steels. // Acta Mater. 2015. V.97. P.316–324

76. *Vaks V.G., Khromov K.Yu.* On the theory of austenite-cementite phase equilibria in steels // JETP. 2008. V.106. №2. P.265–279

77. *Zhang X., Hickel T., Rogal J., Fähler S., Drautz R., Neugebauer J.* Structural transformations among austenite, ferrite and cementite in Fe-C alloys: A unified theory based on ab initio simulations //Acta Mater., 2015. V.99. P.281–289

78. *Aaronson H.I.* Atomic machanisms of diffusional nucleation and growth and comparisons with their counterparts in shear transformations // Metall. Trans. A. 1993. 24A. 241–276

79. *Ali A., Bhadeshia H.K.D.H.* Nucleation of Widmanstätten Ferrite. // Mater.Sci.Technol. 1990. V.6. P.781–784

80. *Yamanaka A., Takaki T., Tomita Y.* Phase-Field Simulation of Austenite to Ferrite Transformation and Widmanstätten Ferrite Formation in Fe-C Alloy. // Materials Transactions. 2006. V.47. №11. P.2725–2731

81. *Chen L.Q., Khachaturyan A.G.* Dynamics of simultaneous ordering and phase-separation and effect of long-range. // PRL. 1993. V.70. P.1477–1480

82. *Allen S.M., Cahn J.W.* Mechanisms of Phase Transformations within the Miscibility Gap of Fe-Rich Fe-Al Alloys // Acta Metall. 1976. V.24. P.425–437

83. *Bray A.J.* Theory of phase-ordering kinetics. //Advances in Physics. 1994. V.43. №3. P.357–459

84. *Falk F.* Model free-energy, mechanics and thermodynamics of shape-memory alloys // Acta Metall. 1980. V.28. P.1773–1780

85. *Onuki A.* Pretransitional Effects at Structural Phase Transitions // J. Phys. Soc. Jpn. 1999. V.68. P.5–8

86. *Kartha S., Krumhansl J.A., Sethna J.P., Wickham L.K.* Disorder-driven pretransitional tweed pattern in martensitic transformations // Phys. Rev. B. 1995. V.52. №2. P. 803–822

87. *Baus M., Lovett R.* Generalization of the stress tensor to nonuniform fluids and solids and its relation to Saint-Venant's strain compatibility conditions // Phys.Rev. Lett. 1990. V.65. №14. P.1781–1783

88. *Khachaturyan A.G.* Theory of Structural Transformations in Solids. New York: Dover, 2008. 592 pp.

89. *De'Bell K., MacIsaac A.B., Whitehead J.P.* Dipolar effects in magnetic thin films and quasi-two-dimensional systems // Rev. Mod. Phys. 2000. V.72. P.225–257

90. *Schmalian J., Wolynes P.G.* Stripe Glasses: Self-Generated Randomness in a Uniformly Frustrated System // Phys. Rev. Lett. 2000. V.85. P.836–839

91. *Jagla E.A.* Numerical simulations of two-dimensional magnetic domain patterns // Phys. Rev. E. 2004. V.**70**. №4. 046204 (7 pp)

92. *Emery V.J., Kivelson S.A.* Frustrated electronic phase separation and high-temperature superconductors // Physica C. 1993. V.209. P.597–621

93. *Kivelson D., Kivelson S.A., Zhao X., Nussinov Z., Tarjus G.* Statistical Mechanics and its Applications // Physica A. 1995. V.219. P.27–38

94. *Nussinov Z., Rudnick J., Kivelson S.A., Chayes L.N.* Avoided Critical Behavior in O(n) Systems // Phys. Rev. Lett. 1999. V.83. №3. P.472–475

95. *Prudkovskii P.A., Rubtsov A.N., Katsnelson M.I.* Topological defects, pattern evolution, and hysteresis in thin magnetic films // Europhys. Lett. 2006. V.73. P.104–109

96. *Razumov I.K., Gornostyrev Yu.N., Katsnelson M.I.* Intrinsic nanoscale inhomogeneity in ordering systems due to elastic-mediated interactions // Europhys. Lett. 2007. V.80. 66001 (5 pp)

97. *Bručas R., Hafermann H., Katsnelson M.I., Soroka I.L., Eriksson O., Hjörvarsson B.* Magnetization and domain structure of bcc $Fe_{81}Ni_{19}/Co(001)$ // Phys. Rev. B. 2004. 69. №6. 064411 (11 pp)

98. *Landau L.D., Lifschitz E.M.* Theory of Elasticity. 3rd Edition. Oxford: Pergamon, 1986. 195 pp

99. *Barsch G.R., Krumhansl J.A.* Nonlinear and nonlocal continuum model of transformation precursors in martensites. // Metall. Trans. A. 1988. V.19. P.761–775

100. *Jiang E., Carter E.A.* Carbon dissolution and diffusion in ferrite and austenite from first principles // Phys. Rev. B. 2003. V.67. 214103 (11 pp)

101. *Lobo J.A., Geiger G.H.* Thermodynamics of carbon in austenite and Fe-Mo austenite // Met. Trans. A. 1976. V.7. №8. P.1359–1364

102. *Mogutnov B.M., Tomilin I.A., Shvartsman L.A.* Thermodynamics of carbon-iron alloys. Moscow.: Metallurgy, 1972. 328 p. [In Russian]

103. *Kurz W., Fisher D.J.* Fundamentals of Solidification, 3rd. ed., Trans Tech Publications, Aedermannsdorf, Switzerland, 1992. 293 pp.

104. *Hecht U., Granasy L., Pusztai T. et al.* Multiphase solidification in multicomponent alloys // Mater. Sci. Eng. R 2004. V.46. P.1–49

105. *Folch R., Plapp M.* Quantitative phase-field modeling of two-phase growth // Phys. Rev. E. 2005. V.72. 011602 (27 pp)

106. *Nestler B., Wheeler A.A.* A multi-phase-field model of eutectic and peritectic alloys: numerical simulation of growth structures // Physica D. 2000. V.138. P.114–133

107. *Boettinger W.J., Warren J.A., Beckermann C., Karma A.* Phase-field simulation of solidification // Annu. Rev. Mater. 2002. V.32. P.163–194

108. *Elder K.R., Drolet F., Kosterlitz J.M., Grant M.* Stochastic Eutectic Growth. // PRL. 1994. V.72. №5. P.677–680

109. *Drolet F., Elder K.R., Grant M., Kosterlitz J.M.* Phase-field modeling of eutectic growth. // Phys. Rev. E. 2000. V.61. №6. P.6705–6720

110. *Greenwood M., Ofori-Opoku N., Rottler J., Provatas N.* Modeling structural transformations in binary alloys

with phase field crystals. // Phys. Rev. B. 2011. V.84. 0641104 (10 pp.)

111. *Kundin J., Raabe D., Emmerich H.* A phase-field model for incoherent martensitic transformations including plastic accomodation processes in the austenite. // J. Mech. Phys. Solids. 2011. V.59. P.2082–2102

112. *Malik A., Yeddu H.K., Amberg G., Borgenstam A., Ågren J.* Three dimensional elasto-plastic phase field simulation of martensitic transformation in polycrystal // Mat.Sci.Eng. A. 2012. V.556. P.221–232

113. *Yeddu H.K., Borgenstam A., Ågren J.* Stress-assisted martensitic transformations in steels: A 3-D phase-field study // Acta Mater. 2013. V.61. P.2595–2606

114. *Levitas V.I., Javanbakht M.* Interaction between phase transformations and dislocations at the nanoscale. // J. Mech. Phys. Solids. 2015. V.82. P.287–319

115. *Roitburd A.L., Temkin D.E.* Plastic deformation and thermodynamic hysteresis at phase transformation in solids // Sov. Phys. Solid State. 1986. V.28. P.432–443

116. *Kaganova I.M., Roitburd A.L.* Effect of plastic deformation on the equilibrium shape of a new phase inclusion and thermodynamic hysteresis // Sov. Phys. Solid State. 1989. V.31. P.545–550

117. *Levitas V.I., Idesman A.V., Olson G.B., Stein E.* Numerical Modeling of Martensite Growth in Elastoplastic Material // Philos. Mag. A. 2002. V.82. P.429–462

118. *Fisher F.D., Reisner G.* A criterion for the martensitic transformation of a microregion in an elastic-plastic material // Acta Mater. 1998. V.46. №6. P.2095–2102

119. *Smart J.S.* Effective field theories of magnetism. Saunders, 1968. 188 pp.

120. *Körmann F., Dick A., Grabowski B., Hallstedt B., Hickel T., Neugebauer J.* Free energy of bcc iron: Integrated ab initio derivation of vibrational, electronic, and magnetic contributions // Phys. Rev. B. 2008. V.78. №3. 033102 (4 pp)

121. *Ziman J.M.* Models of Disorder. Cambridge University Press, 1979. 542 pp.

122. *Smith W.F., Hashemi J.* Foundations of Materials Science and Engineering. McGraw-Hill, Allas, USA, ed.4, 2005. 1056 pp.

123. *Lavrentiev M.Yu., Nguyen-Manh D., Dudarev S.L.* Magnetic cluster expansion model for bcc-fcc transitions in Fe and Fe-Cr alloys // Phys. Rev. B. 2010. V.81. №18. 184202 (6 pp)

124. *Chen Q., Sundman B.* Modeling of Thermodynamic Properties for Bcc, Fcc, Liquid, and Amourphous Iron // Journal of Phase Equilibria. 2001. V.22. №6. P.631–643

125. *Darken J.S., Gurry R.W.* Free Energy of Formation of Cementite and the Solubility of Cementite in Austenite // Trans. AIME. 1951. V.191. P.1015–1018

126. *Dick A., Körmann F., Hickel T., Neugebauer J.* Ab initio based determination of thermodynamic properties of cementite including vibronic, magnetic, and electronic excitations // Phys. Rev. B. 2011. V.84. 125101 ( pp.)

127. *Battezzati L., Baricco M., Curiotto S.* Non-stoichiometric cementite by rapid solidification of cast iron. // Acta Mater. 2005. V.53. P.1849–1856

128. *Mehrer H.* (Ed.): Landolt-Börnstein, Numerical Data and Functional Relationships in Science and Technology, New Series, Group III: Crystal and Solid State Physics, V. 26, Diffusion in Metals and Alloys, Springer-Verlag, Berlin, 1990. 747 pp.

129. *Ozturk B.* The diffusion coefficient of carbon in cementite, $Fe_3C$, at $450^0C$. //Solid State Ionics. 1984. V.12. P.145–151

130. *Hultgren A.* Isothermal transformation of austenite. // Trans. ASM. 1947. V.39. P.915–1005.

131. *Aranda M.M., Rementeria R., Capdevila C., Hackenberg R.E.* Can Pearlite form Outside of the Hultgren Extrapolation of the $A_{e3}$ and $A_{cm}$ Phase Boundaries? // Met. Mat. Trans. A. 2016. V.47. P.649–660

132. *Christian J.W.* Thermodynamics and kinetics of martensite. // In: Intern. Conf. on Martensitic Transformations, ICOMAT' 79. Ed. by Olson G.B., Cohen M. Boston, USA. P.220–234

133. *Lobodyuk V.A., Estrin E.I.* Isothermal martensitic transformations. // Physics-Uspekhi. 2005. V.48. N7. P.713–732.

134. *Manna I., Pabi S.K, Gust W.* Discontinuous Reaction in Solids // Inter. Mater. Rev. 2001. V.46. P.53–91

135. *Bensaada S., Mazouz H., Bouziane M.T.* Discontinuous Precipitation and Dissolution in Cu-4.6at.% In Alloy under Effect of Plastic Deformation and the Temperature // Mat. Sci. App. 2011. V.2. P.1471–1479

136. *Hornbogen E.* Systematics of cellular precipitation reactions. // Met. Mat. Trans. B. 1972. V.3. №11. P.2717–2727

137. *Ramanarayan H., Abinandanan T.* Grain boundary effects on spinodal decomposition. II Discontinuous microstructures // Acta Mat. 2004. V.52. P.921–930

138. *Razumov I.K.* The Simulation of the Growth of Colonies in the Spinodal Decomposition of Metastable Phases. // Russian Journal of Physical Chemistry A. 2009. V.83. №10. P.1682–1688

139. *Razumov I.K.* Influence of lattice relaxation on the kinetics of spinodal decomposition of solid solutions. //JEPTER. 2009. V.82. №4. P.635–641

140. *Smith C. S.* Microstructure. // Trans. Am. Soc. Metals. 1953. V.45. P.533–575

141. *Sadovskii V.D., Rodigin N.M., Smirnov L.V., Filonchik G.M., Fakidov I.G.* On magnetic field effect on martensitic transformation in steel // Fiz.Metal.Metalloved. 1961. V.12. P.302–304

142. *Fokina E.A.*, Zawadskii E.A. The effect of magnetic field on martensitic transformation in steel // Fiz. Metal. Metalloved. 1963. V.16. №2. P.311-313

143. *Sadovskii V.D., Smirnov L.V., Fokina E.A.* Steel Quenching in Magnetic Field // Fiz. Metal. Metalloved. 1967. V.24. №5. P.918–939

144. *Malinen P.A., Sadovskii V.D., Smirnov L.V., Fokina E.A.* On principles of the pulsed magnetic field effect

on martensitic transformations in steel and alloys // Fiz. Met. Metalloved. 1967. V.23. P. 535–542

145. *Krivoglaz M.A., Sadovskii V.D., Smirnov L.V., Fokina E.A.* Steel Quenching in Magnetic Field. Moscow: Nauka, 1977 [In Russian]

146. *Estrin E.I.* Magnetic field effect on martensitic transformation // Fiz. Met. Metalloved. 1965. V.19. P. 929-932

147. *Shimotomai M., Maruta K.* Aligned two-phase structures in Fe-C alloys. // Scripta Mater. 2000. V.42. №5. P.499–503

148. *Shimotomai M., Maruta K., Mine K., Matsui M.* Formation of aligned two-phase microstructures by applying a magnetic field during the austenite to ferrite transformation in steels // Acta Mater. 2003. V.51. №10. P.2921–2932

149. *Zhang Y.D., Gey N., He C.S., Zhao X., Zuo L., Esling C.* High temperature tempering behaviors in a structural steel under high magnetic field // Acta Mater. 2004. V.52. №12. P.3467–3474

150. *Zhang Y.D., Esling C., Zhao X., Zuo L.* Solid State Phase Transformations under High Magnetic Fields in a Medium Carbon Steel // Mat.Sci.Forum. 2005. V.495–497. P.1131–1140

151. *Krivoglaz M.A., Sadovskii V.D.* On strong magnetic field effect on phase transformations // Fiz. Met. Metalloved. 1964. V.18. №4. P.502–505

152. *Abrikosov I.A., Ponomareva A.V, Steneteg P., Barannikova S.A., Alling B.* Recent progress in simulations of the paramagnetic state of magnetic materials. // Current Opinion in Solid State and Materials Science. 2016. V.20. P.85–106.